%% file: main.tex
\documentclass[fleqn,usenatbib,useAMS]{mnras}

\usepackage{newtxtext,newtxmath}
\usepackage[T1]{fontenc}

\usepackage{graphicx}
\usepackage{pdflscape}
\usepackage{rotating}
\usepackage{multicol}
\usepackage{ae,aecompl}
\usepackage{tabularx}
\usepackage{multicol}

\title[Detecting Pulsars with Neural Networks]{Detecting Pulsars with Neural Networks: A Proof of Concept}
\author[L.\ K\"unkel et al.]{Lars K\"unkel$^{1}$,
Rajat M.\ Thomas$^{2}$,
and Joris P.~W.\ Verbiest$^{1,3}$
\\$^{1}$ Fakult\"at f\"ur Physik, Universit\"at Bielefeld, Postfach 100131,
33501 Bielefeld, Germany\\
$^{2}$ Netherlands Institute for Neuroscience, Royal Netherlands Academy of Arts and Sciences, Amsterdam, the Netherlands\\
$^{3}$ Max-Planck-Institut f\"ur Radioastronomie, Auf dem H\"ugel 69, 53121 Bonn, Germany}

\date{Accepted XXX.  Received YYY; in original form ZZZ}

\pubyear{2021}

\begin{document}
\label{firstpage}
\pagerange{\pageref{firstpage}--\pageref{lastpage}}
\maketitle

\begin{abstract}
  \input{abstract}
\end{abstract}

\section{Introduction}
    \input{introduction_3}

\section{Related Work and Concepts}
We start here by describing traditional approaches to pulsar searches followed by
examples of the application of machine learning techniques in pulsar astronomy.
We then provide a brief primer on the neural network concepts that are relevant to our proposed model.

\subsection{Common Pulsar Survey Techniques}
    \label{sec:survey}
    \input{related_work_survey}

\subsection{Machine Learning in Pulsar Astronomy}
    \input{related_work_astroml}

\subsection{Neural Networks}
\input{related_work_nn}

\section{Methods}
We implemented the neural networks using \textsc{PyTorch}\footnote{An open-source framework for deep learning \url{https://pytorch.org}. The results presented in this work were trained using \textsc{pytorch} 1.1.0.} \citep{NEURIPS2019_9015}.
In this Section we will describe the architecture of the neural network which is shown in Figure \ref{fig:architecture}, the training objectives which are used to train the network, the flavours of neural networks we use and the training procedure we use to train the network.
\subsection{Neural Network Architecture}
    \input{architecture}

\subsection{Trained Models}
    \input{models}
    
\subsection{Performance Metrics}
    \input{perf_metrics}
    
\subsection{Training Procedure}
    \input{training}

\section{Data}
\label{sec:data}
In this section we will describe the data of the PALFA survey which we analyze in this work and the simulations that are used to to train the neural network. Table \ref{tab:set_size} summarizes the size of the used data sets.

\input{tables/set_size}
\input{tables/psr_table_pred.tex}
\subsection{Observations}
    \input{observations}
\subsection{Simulations}
    \input{simulations}

\section{Results}
In this section we describe the performance of the neural networks. First we assess the quality of the dedispersed output using the FFA (Section \ref{sec:ffa_perf}) and afterwards we judge the classification performance of our model on simulated (Section \ref{sec:perf_fake}) and real (Section \ref{sec:perf_real}) pulsars. In Section \ref{sec:rfi} we discuss how well the model copes with RFI.
\subsection{FFA Performance}
    \input{ffa_performance}
\subsection{Classifier Performance on Simulated Data}
    \input{class_performance_fake}
\subsection{Classifier Performance on Real Data}
    \input{class_performance_real}

\subsection{Influence of RFI}
    \input{rfi}

\section{Applicability to Pulsar Surveys}
    \input{application}

\section{Conclusion}
    \input{conclusion}

\section*{Acknowledgements}
    \input{acknowledgements}
    
\section*{Data Availability}
    The data underlying this article are publicly available observations of the PALFA survey.
    Details of the survey can be found at \url{https://www.naic.edu/alfa/pulsar/}.
\bibliographystyle{mnras}
\bibliography{./references}

\bsp
\label{lastpage}
\end{document}

%% file: abstract.tex
Pulsar searches are computationally demanding efforts to discover
dispersed periodic signals in time- and frequency-resolved data from radio
telescopes. The complexity and computational expense of simultaneously
determining the frequency-dependent delay (dispersion) and the
periodicity of the signal is further exacerbated by the presence of
various types of radio-frequency interference (RFI) and observing-system effects. New observing systems with wider bandwidths, higher bit
rates and greater overall sensitivity (also to RFI) further enhance these
challenges. We present a novel approach to the analysis of pulsar
search data. Specifically, we
present a neural-network-based pipeline that efficiently suppresses a wide
range of RFI signals and instrumental instabilities and furthermore
corrects for (\textit{a priori} unknown) interstellar dispersion. After initial training of the
network, our analysis can be run in real time on a standard desktop computer with a commonly available, consumer-grade GPU. 
We 
complement our neural network with standard algorithms for
periodicity searches. In particular with the Fast Fourier Transform
(FFT) and the Fast Folding Algorithm (FFA) and demonstrate that with
these straightforward extensions, our method is capable of
identifying even faint pulsars, while maintaining an extremely low
number of false positives. We furthermore apply our analysis to a subset 
of the PALFA survey and demonstrate that in most
cases the automated dispersion removal of our network produces a time series of similar quality as dedispersing using the actual dispersion measure of the pulsar
in question. 
On our test data we are able to make predictions whether a pulsar is present in the data or not 200 times faster than real time.

%% file: introduction_3.tex
Pulsars are rotating neutron stars which can be observed in radio data as regular,  dispersed pulses with a pulsar-specific periodicity. 
The majority of
pulsars are too faint to be detectable as single
pulses \citep{2004hpa..book.....L}, which necessitates the use of
techniques 
such as the {\it Fast Fourier Transform} (FFT) or the {\it Fast Folding Algorithm}
\citep[FFA,][]{1969A&A.....2..280B}, which use the inherent periodicity
of the pulsar signal to allow a detection. 
During pulsar searches the {\it dispersion measure} (DM) of the pulsars 
which defines the frequency-dependent delay due to propagation through the interstellar medium of the pulsar's pulses is not known a priori.
To circumvent this problem many trials at different DM values are computed
and are subsequently analysed.
Due to the faintness of the pulsar signals traditional survey pipelines create
an immense number of pulsar candidates  
which are mostly the result {\it radio-frequency interference} (RFI) and 
other forms of noise such as noise created by the observing-system.
The large number of pulsar candidates subsequently have to be classified.
Manual inspection of such large numbers of (mostly false)
candidates is extremely labour-intensive \citep{2020AmJPh..88...31B} and demonstrably likely to miss out on real pulsars. Automated classification and ranking schemes have been introduced to alleviate this challenge \citep{2016MNRAS.459.1104L,lsj+13}. Such classification pipelines use the search-derived parameters to build diagnostic plots and quantities, which are then used to discern pulsars from non-pulsars.

In this paper, unlike in previous approaches, we attempt to build an {\it end-to-end} pipeline using convolutional neural networks to detect faint pulsars in noisy environments by (i) automatically suppressing RFI and instrumental instabilities, (ii) correcting dispersive effects in the time series, and (iii) assessing the likely presence of periodic signals. The developed software is available at \url{https://github.com/larskuenkel/DeepPulsarNet}.
Section 2 gives an introduction to pulsar survey techniques and relevant aspects of neural networks.
In Section 3 we describe our neural network architecture and detail the signal we approximate using the neural network.
In Section 4 we introduce both the simulated and real data that are used for training and validation of the neural network model.
In Section 5 we analyze the performance of the neural network.
Section 6 outlines how our method could be used in future pulsar surveys.
We conclude in Section 7.

%% file: related_work_survey.tex
An extensive description of
the steps involved in modern-day pulsar surveys and the details
involved in such analyses, can be found in \citet{bar13}. Here we
give a brief overview of the main points relevant to our work.

Pulsar survey observations are analyzed as 2D arrays of intensity as a function of time and observing frequency. The first step in the 
analysis pipeline is the mitigation of RFI.
Following that, the effect of interstellar dispersion is removed.
The presence of free electrons along the line of sight to the pulsar creates
a frequency-dependent delay of the arrival time of the pulses.
This delay between two frequency channels with observing frequencies $f_1$ and $f_2$ in MHz is given by \citep{2004hpa..book.....L}:

\begin{equation}
\label{eq:dm_delay}
\Delta t \simeq 4.15 \times 10^6 \text{ms} \times (f_1^{-2} \times f_2^{-2}) \times \text{DM},
\end{equation}
where the dispersion measure DM in $\text{cm}^{-3}\,\text{pc}$ is defined by the electron number density $n_e$  according to
\begin{equation}
\label{eq:dm}
\text{DM} = \int_{0}^{d}n_e\text{d}l.
\end{equation}

To mitigate for dispersion at a specific DM the frequency channels in an observation are shifted based on Equation \ref{eq:dm_delay} and the different channels are summed afterwards resulting in one dedispersed 1D time series.
In a survey setting the DM of the pulsars which stays relatively stable over time
is not known prior to detection which is why
observations are dedispersed at a range of different DMs.
This range is chosen to minimize the broadening that is
introduced to the pulses by not dedispersing them at the true DM of
the pulsar. Depending on the survey specification this 
can result in tens
to thousands of DM trials being created for every
observation \citep{2016MNRAS.459.1104L}.

Most pulsars are not bright enough to be detected as repeated
individual pulses above the background noise \citep{2004hpa..book.....L}, but 
instead require the use of techniques which utilise the periodicity
of the pulsar signal to improve the signal-to-noise ratio (S/N). Historically most pulsars have been discovered using
FFT based techniques but recently interest in the use of
the FFA has been renewed
\citep{2017MNRAS.468.1994C,2018ApJ...861...44P,2020MNRAS.497.4654M}.
When using the FFT to search for pulsar pulses, narrower pulses result in more power being distributed into higher harmonics of the pulsar's spin frequency. To alleviate this problem in Fourier transform based
pipelines the different harmonics are added by {\it harmonic summing} \citep{1969Natur.221..816T}.
The FFT or the FFA are then investigated for strong peaks which might
constitute a pulsar.
By comparing the peaks of multiple DM trials
unlikely candidates can be sifted out.
The most likely pulsar candidates are folded in time to create a richer representation of the pulsar. Along with the folded pulse profile, various diagnostic plots which are instrumental to determining the authenticity of the pulsar are also generated. The classification of whether the signal is a true pulsar or not was previously done by human experts but has in recent times increasingly been automated.

Many pulsars exist as binary systems with another object which leads to a
modulation of the observed pulse frequency. The Fourier transform is
less sensitive to these modulated frequencies. Different methods of
acceleration search have been developed to increase the sensitivity of
the FFT to such modulated signals \citep[e.g.][]{2002AJ....124.1788R}. These methods heavily
increase the computational requirements of the survey and also
increase the number of false positive candidates.

%% file: related_work_astroml.tex
\label{sec:astroml}

The application of machine learning techniques to pulsar detection in recent years has focused on automating the classification of pre-selected {\it pulsar candidates}.
Modern pulsar surveys with their massive amounts of data require the development of automated systems to handle the large number of expected candidates \citep{2016MNRAS.459.1104L}.
Since the largest amount of pulsars are found using periodicity-based searching techniques most classification schemes use folded candidates as the input for the classification.
The automated systems to classify folded candidates have been using various machine learning techniques such as decision trees, support vector machines and neural networks \citep[see e.g.][]{2014ApJ...781..117Z,2018MNRAS.474.4571T,2019SCPMA..6259507W}

Folded pulsar-candidate plots try to summarize the observational characteristics of the candidate. Information that has been discarded during the initial search process like the temporal and spectral persistence of the pulsar signal are shown in such typical diagnostic plots \citep[][ Fig 1]{2018MNRAS.474.4571T}. 
Although the features in the candidate plots are designed to investigate the characteristics of pulsars, building an automated system to discern a pulsar candidate from a spurious candidate resulting from noise or RFI is daunting due to the sheer number of false pulsar candidates \citep{2016MNRAS.459.1104L}.

While the discussed folded candidates are the result of periodicity based searching techniques, candidates resulting from single pulse searches have been classified in \citet{2018MNRAS.480.3302P}.
In their work single pulses are detected, clustered and the resulting clusters are classified using various supervised machine learning techniques. Their best classifier was a random forest classifier which used {\it SMOTE} (Synthetic Minority Oversampling TEchnique) treatment to better treat the class imbalance in their data set.
Using single-pulse-based searches can help finding signals which are not easily found when using periodicity-based techniques such as sporadic pulsars, FRBs or in very short observations. 

Finding FRBs in radio data using machine learning techniques has been done closer to the raw data than finding pulsars since FRB searches only require a small window of the observation.
Various works have been published that build systems which are largely based on convolutional neural networks (CNNs, see Section \ref{sec:nn_basics}) to detect FRBs \citep{2018AJ....156..256C, 2018ApJ...866..149Z, 2020MNRAS.497.1661A}.
These neural networks use different inputs. \citet{2018ApJ...866..149Z} uses segments of 256 time samples of the raw observation as input to detect single pulses of the repeating FRB 12110. 
They trained a neural network to be receptive to pulses from a wide range of DMs. They trained their network on pulses with a DM between 200 and 2000.
The systems in \citet{2020MNRAS.497.1661A} and \citet{2018AJ....156..256C} require a dedispersed input instead of the raw observation. 
\citet{2020MNRAS.497.1661A} use a segment of the dedispersed frequency-time spectrogram and the DM-time array, which has been averaged along the frequency axis, as the input to the neural network, \citet{2018AJ....156..256C} use the dedispersed frequency-time intensity array, the pulse profile, the DM-time array and the multibeam detection S/N as input.

\citet{2018AJ....156..256C} discuss the theoretical possibility of a forward pass through a neural network replacing brute-force dedispersion. 
They suggest that while a pass through a relatively small convolutional neural network could be faster than brute-force dedispersion, the CNN may not reach the level of statistical optimality without increasing the network size to a size where the gain in speed is lost. 
However \citet{2018ApJ...866..149Z} have shown that neural networks are capable of detecting dispersed pulses. They conclude that their approach of detecting FRBs with neural networks in spectral-temporal data shows potential advantages in both sensitivity and computational speed over dedispersion pipelines.

%% file: related_work_nn.tex
\label{sec:nn_basics}
Our end-to-end approach for the identification of pulsars is based on neural networks. Specially, we utilize a class of neural networks referred to as CNN. In this section, we briefly summarize the principles behind neural networks with emphasis on CNN. A deeper explanation of neural networks can be found in \citet{franoischollet2017learning} for example.

\paragraph*{Neural Network Basics}
A neural network in essence is an approximation of a universal function that maps input $\mathbf{x} \in \mathbb{R}^n $ to output $\mathbf{y} \in \mathbb{R}^m$. Neural networks are composed of so-called layers that perform a weighted average of their inputs typically followed by a non-linear {\it activation} or transformation. $L$ such layers with the output of one being the input to the other compose what is called a Multi-layer perceptron (MLP). Mathematically, operations of a layer in an MLP can be written as:
\begin{equation}
\label{eq:mlp}
\mathbf{x}_{l+1} = \sigma(\mathbf{W}_l \mathbf{x}_l + \mathbf{b}_l),
\end{equation}
where $\mathbf{x}_l$ are the inputs to layer $l+1$, with $\mathbf{x}_0$ being the input $\mathbf{x}$ and final $\mathbf{x}_L$ being the output values $\mathbf{y}$. $\sigma$ is the non-linear activation function. 
The parameters of the affine transform often referred to as weights $\mathbf{W}_l$ and biases $\mathbf{b}_l$ are what powers neural networks.
The parameters $\mathbf{W}_l$ and  $\mathbf{b}_l$ are initialized randomly and over the course of {\it training} the neural network they morph into data-driven detectors of features that are pertinent to the task, which in our case for example is the classification of pulsars.
In an MLP the user defines the number of output values
in each layers which results in the dimensionality of $\mathbf{W}_l \in \mathbb{R}^{o \times p}$.
In the first layer $p$ is given by the input size $n$ and in the last layer $o$ is given by the output size $m$.
Without the activation functions an MLP would only learn a linear transformation
of the input values
with an added bias term.
The addition of the activation function allows the network to identify
more complex relations in the data.

The key to making the network give accurate predictions is training the network using gradient descent.
This is done by comparing the prediction of the network and the expected output using a {\it loss function} and altering the value of the weights and biases using the gradient of the loss function.
In order to determine the prediction, a trained network will transform the input features $\mathbf{x}$ into more useful representations $\mathbf{x}_l$.
Using \textit{deep} neural networks with many layers has been proven to achieve good predictions on a vast variety of problems.

The number of parameters in a MLP scales with the number of input parameters. This makes it not ideal when handling a large amount of input values like the very long time series in pulsar survey data.

\paragraph*{Convolutional Neural Networks}
Convolutional neural networks are a very potent tool to apply to data where translational invariance can be exploited, like images \citep{fawaz2019deep} or sequences that have a temporal order \citep{2018arXiv180301271B}.
In a convolutional layer the output is calculated by convolving small kernels with the input values followed by an activation function.
By using these small kernels with trainable weights the network is able to create useful representations of features in the input regardless of the position of the feature in the input.
The output size of a convolutional layer depends on the stride of the convolution. 
The output size may be reduced through the use of pooling layers that replace the value of adjacent values by the maximum or the average of these pixels.
Many different architectures have been built on the basis of convolutional layers and many different aspects of the network can be altered to make the network more effective for a specific problem.
One of these aspects is the usage of dilated convolutions \citep{2018arXiv180301271B}. These convolutions do not convolve adjacent pixels with the kernels but instead uses spaced out pixels for the convolutions.
This allows the network to reach a high receptive field even without pooling layers.

Another innovation in deep neural networks has been the introduction of {\it residual blocks}. In these networks called ResNets \citep{He_2016_CVPR} the output of one layer is added to the input that went in to that layer before feeding the result to the next convolutional layer.
This allows training in very deep networks where otherwise vanishing gradients would be a problem.
A network that uses residual blocks can learn to skip layers if they do not help the performance of the network.

Neural networks tend to perform better on data that have been seen during training than on unseen data.
{\it Regularization} is the process of mitigating this problem. A commonly used regularization technique is {\it dropout} where each input value has a set chance of being set to zero. Dropout layers can be used throughout the network.

{\it Normalization layers} will help the training to converge faster and can also help regularizing the network. 
These layers adjust the input values to have a suitable distribution for the neural network.
Most commonly they transform the input to have zero mean and a standard deviation of one.

How exactly the weights of the neural network are updated during training is defined by an {\it optimizer}. In this work we use the Adam optimizer \citep{2014arXiv1412.6980K} included in \textsc{PyTorch}.

%% file: architecture.tex
\label{sec:arch}

\subsubsection{Outline of the Neural Network}
The goal of the network is classifying whether a pulsar search observation contains a pulsar or not. 
In order to do this we use a neural network with two parts:  dedisperision and classification. 
The dedispersion part is described in detail in Section \ref{dedis_arch}. Briefly, it creates 
1D time series which contain all pulses that the network sees in the data, similar to the results of conventional dedispersion
methods.

In conventional dedispersion, given a specific DM value, different frequency channels are combined (averaged) after applying a frequency-dependent time delay associated with that DM. This implies that every time point in the averaged time-series is a function of a narrow temporal window at each frequency. In contrast, for our neural network, a single output time point is a function of several seconds of temporal window across all frequencies --- a property that would allow it to be sensitive to a range of DMs in one step while classical techniques require multiple DM trials.

The classifying part of the network classifies the 1D time series and is described in Section \ref{class_arch}. Instead of using a neural network to classify the resulting time series, the time series can also be investigated for pulsar candidates using standard pulsar search techniques that are normally applied to dedispersed time series.
The classifiers in this work only uses the strongest signal in the dedispersed time series which assumes that there is only one pulsar at most in an observation.
For pulsar surveys, which utilise a large field of view and possibly contain a larger number of pulsars in a single observations, the classifiers could be adapted to produce a candidate list instead of a classification of the whole observation.

\begin{figure*}
 \includegraphics[width=1.2\columnwidth]{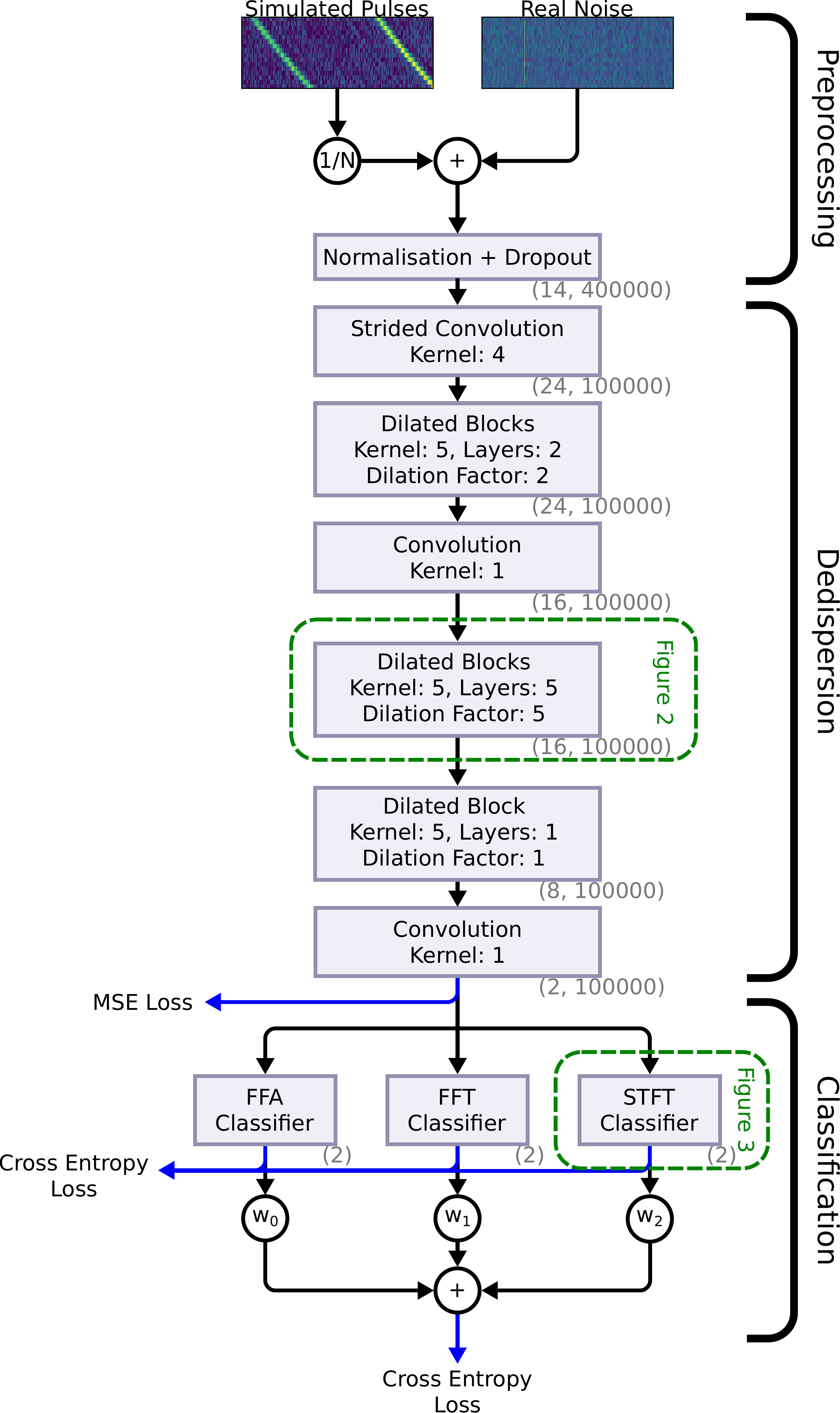}
 \centering
 \caption{Architecture of the complete neural network. 
 The numbers at the bottom right of each block indicate the data volume of the output of the block.
 In the dedispersion network we use 1D convolution which means that the first entry of the data
 volume is a chosen number of channels and the second entry is the number of time steps.
 The blue lines indicate at which stages the various losses and the gradients which are used to train the network are calculated. The structure of one layer of the dilation blocks is shown in Figure \ref{fig:single_block}. The structure of a STFT classifier is shown in Figure \ref{fig:stft_class}.}
 \label{fig:architecture}
\end{figure*}

\subsubsection{Preprocessing and Regularization}\label{sec:preprocess}

The input to the network are filterbank files which can contain simulated or real pulsars. The details about using simulated pulsars are included in Section \ref{sec:simulations}.
The observations are loaded using  \textsc{sigpyproc}\footnote{\hypertarget{https://github.com/ewanbarr/sigpyproc}{https://github.com/ewanbarr/sigpyproc}}.
As part of the preprocessing channels corresponding to the highest and lowest observing frequency are eliminated as they mostly contain noise.
To normalize the input values we subtract the mean of all input values from each value and divide all values by the standard deviation of the input values.
To regularize the network we use a dropout layer (see Section \ref{sec:nn_basics}) with a dropout value of 0.1 after normalization which means that during training 10\% of the input data is zeroed.

\subsubsection{Dedispersion Architecture}\label{dedis_arch} 

\begin{figure}
 \centering
 \includegraphics[scale=0.45]{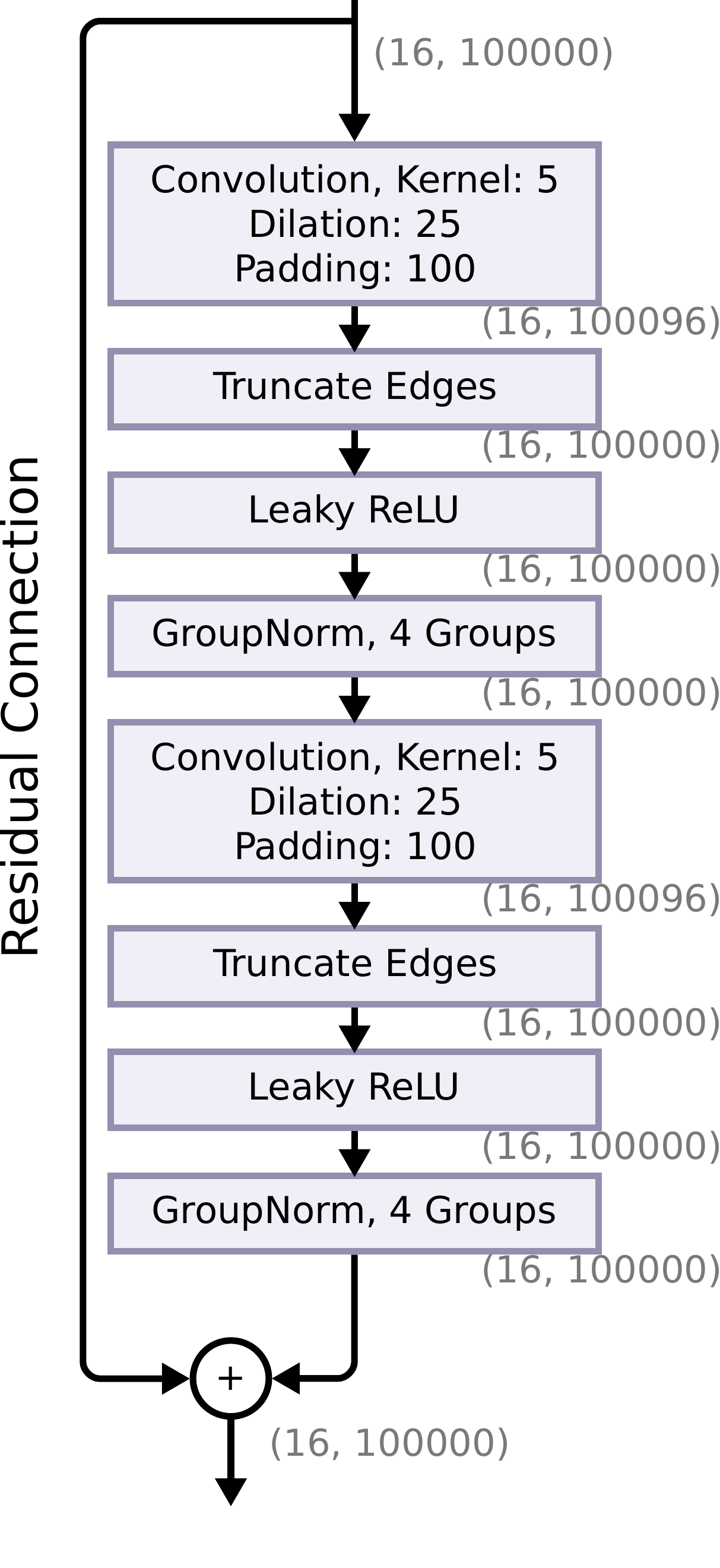}
 \caption{Exemplary architecture of the third layer of the second dilated block which is indicated in Figure \ref{fig:architecture}. The whole dilated block has 5 layers. The other layers share the same properties except for the dilation which is increasing with each layer.
The data volume shown in grey is kept constant by choosing the right padding value and truncating the edges.}
 \label{fig:single_block}
\end{figure}

The main part of the network is based on the Temporal Convolutional Networks (TCN) described in \citet{2018arXiv180301271B}.  This network type mainly consists of 1D dilated convolutions where the output of each layer has the same cardinality as the input to that layer. The modifications we made to the canonical TCN in our architecture are shown in Figure \ref{fig:single_block}; we added a GroupNorm layer \citep{2018arXiv180308494W} after every non-linearity which  resulted in faster convergence and the convolutions are non-causal, i.e., the output depends on both past and future values.
Since we are post-processing the data, this is not a problem.

 Given the dimensions of the input data and the size of the network, it is evident that we need a few "information bottlenecks" to tame the memory requirements of the proposed model. In the initial layers therefore, we utilise strided convolutions to effectively downsample our temporal resolution by a factor of four;  thus going from 400,000 time steps at the input to 100,000 after the first convolutional layer.

Following the initial strided convolutions the data are fed through two blocks of TCNs. The first block has a dilation rate of $d=2^L$ \citep[as in][]{2018arXiv180301271B} where L is the depth of the layer, starting with 0, while the dilation rate  in the second block \citep[as in][]{2019arXiv190300695C} is chosen to be $d=w^L$ where $w$ is the kernel size which is 4 in this work. The increased dilation rate in the second block allows us to reach a higher receptive field with fewer layers. Between the two blocks the number of channels is reduced by a convolution with kernel size one, while the number of channels in each block is kept constant.

 After the dilated blocks, a single residual block with a dilation and kernel size of one outputs an estimation of the dedispersed time series.
The last convolution in the dedispersion network can have multiple output channels where each output channel can be seen as a different time series which can be subsequently analysed.

In our network design we create multiple time series as the output of the neural network dedispersion. Because the neural network tries to work out multiple different DMs that might be plausible, we made this deliberate choice to create two different time series as outputs to enable the neural network to separate the low- and high-DM pulsars (details in Section \ref{train_obj}).

\subsubsection{Classifying Architecture}\label{class_arch} 

\begin{figure}
 \centering
 \includegraphics[scale=0.45]{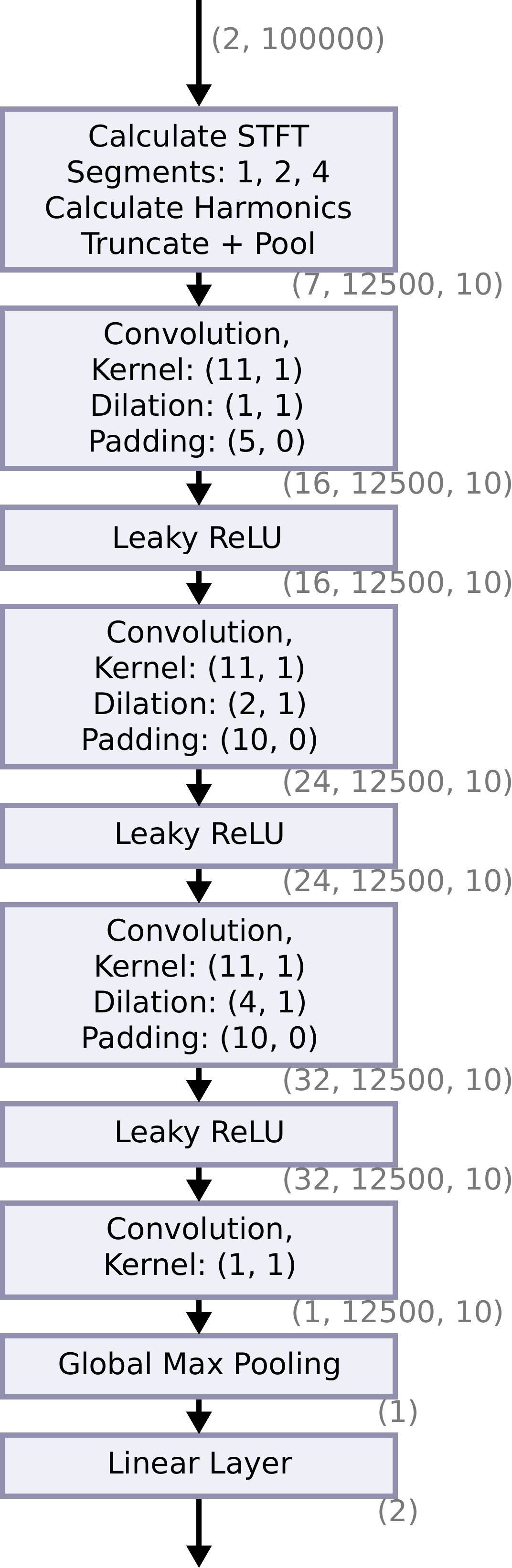}
 \caption{Architecture of the STFT classifier. In this case the dedispersion network has two output channels. The calculation of the STFT results in the FFT of 7 different segments.
These segments show up in the the first dimension of the data volume which is shown in grey after each layer.
 For each input channel five different harmonic combinations are created which results in ten channels which is seen in the last dimension of the data volume. 
 We are using 2D-convolutions here but the kernel size in the last dimension is always one which means that the different harmonic combinations are only combined in the global max pooling layer. The output size two is required by the used cross-entropy loss where each value represents the likelihood of the pulsar either containing a pulsar or not containing a pulsar.}
 \label{fig:stft_class}
\end{figure}

The second part is the classifying part of the network. Since pulsar signals usually are very faint but show periodicity we decided to utilise the 
FFT and the FFA in the classification as in traditional pulsar searches (see Section \ref{sec:survey}). 
The classification is the result of the combination of three individual classifiers; based on FFA, FFT, and the Short-Time Fourier Transform (STFT). Each classifier is trained individually and the results are linearly combined using learnable weights to  deliver the final results.

Utilising the STFT in one of the classifiers is done due to the nature of pulsar signals. Using the STFT allows the network to observe the temporal stability of the pulsar signal when doing the classification which is not done in the FFT and FFA classifiers. While in this study we only look at non-accelerated pulsars, the structure of the classifier allows us in theory to also be more sensitive to some accelerated pulsars than a simple FFT classifier would be.
Calculating the STFT will result in less power of an accelerated pulsar being spread out over multiple FFT bins.
Similar to the stack search described in \citet{2004MNRAS.355..147F}, a neural network could recover an accelerated signal by combining the individual segments of the STFT.

It should be noted that the classification in our neural network works under tighter restrictions than classical pulsar searching pipelines. While usual pipelines create a range of pulsar candidates for a single observation, this classifier tries to classify if the entire observation contains a pulsar or not.

\paragraph*{FFT classifier}
This classifier uses the FFT of the dedispersed time series as input. The spectral power of periodic signals calculated via the FFT is reflected not only at the true periodicity but also in the various harmonics of the period. To mitigate this effect, we implemented the incoherent harmonic summing technique  first introduced in \citet{1969Natur.221..816T}; the magnitude of the FFT output is stretched by various factors and added to the original FFT. In our case, this resulted in five different sequences - the original FFT and four harmonically combined versions up to the second, fourth, eighth and sixteenth harmonics. To keep the noise level identical between sequences, each sequence is divided by $\sqrt{N}$ where $N$ is the number of combined harmonics. 
High frequencies are truncated since high frequency noise sometimes leads to misclassification.
Techniques to reduce the effect of FFT scalloping where the power of a pulsar may be reduced if the pulse period lies between FFT bins have not been employed.

If the dedispersion part of the network has multiple output channels the result of the FFT is concatenated in the same dimension as the harmonic combinations.
After calculating the FFT, truncating high frequencies and summing the harmonics this results in an output shape for a single observation with two output channels in the dedispersion network of:
\begin{equation*}
\label{eq:shape}
(40000, 10) 
= 
\text{(fft frequencies, harmonic sums} \times \text{channels)}.
\end{equation*}

The FFT output is passed through a small convolutional net which outputs a single channel. 
The convolutions are applied over all harmonic combinations separately. 
The single output channel ideally, like in normal pulsar search algorithms, represents a measure of the significance of the peaks in the Fourier transform.
In order to obtain a classification result for the whole observation we subsequently apply global max pooling along the dimension containing the FFT frequencies and the dimension containing the different harmonic combinations. 
This global pooling allows us to find the most significant and pulsar-like signal in the observation and localise the frequency of this signal. 
The amplitude of the output of the pooling layer is used for the classification result of the classifier.
Because the convolutions are applied over the different harmonic combinations separately the final classification result is only the result of the channel which is most likely to contain a pulsar and not the combination of multiple channels.

\paragraph*{STFT Classifier}
The STFT classifier has a similar structure to the FFT classifier. The structure of the classifier can be seen in Figure~\ref{fig:stft_class}. The only difference is that while the FFT based classifier only uses the FFT as input, the STFT classifier also computes two STFTs by splitting up the time series into two and four non-overlapping segments and calculating the FFT for those. 
This results in seven FFTs. In order to process these FFTs with different sizes with a CNN we apply average pooling in order to achieve the same length and frequency resolution in the individual FFTs.
Before feeding them to a convolutional neural network the FFTs are concatenated along the channel dimension which means that the classifying convolutional neural network first trains kernels which combine these different FFTs. 
This allows the convolutional neural network to see how the strength of a signal of a certain frequency develops during the observation.

\paragraph*{FFA classifier}
The FFA classifier computes the FFA of the intermediate time series using the Python library \textsc{riptide}\footnote{\hypertarget{https://github.com/v-morello/riptide}{https://github.com/v-morello/riptide}} \citep{2020MNRAS.497.4654M}. To gain sensitivity to the range of pulsar periods which are present in the test set we compute the FFA in three period ranges using the parameters shown in Table~\ref{tab:ffa_para}. 

We also compute the period-dependent detection threshold of the peak detection algorithm of the riptide library and subtract this threshold from the resulting FFA. This eases detection by removing large-scale variations and reducing the effect of noisy segments of the FFA.
The result is fed into a similar network as for the other classifiers. In the case of the FFA no harmonic pooling is necessary as the result is phase coherent.

The FFA calculation is based on an existing library which does not allow us to propagate the gradients through it and is running on a CPU. In this implementation the calculation of the FFA can increase the time needed for one training loop by a factor of 10. Since the FFA  classifier does not provide useful gradients for the training of the dedispersion network the FFA classifier is only added to the model in the last step of training.

\input{tables/ffa_para.tex}

\subsection{Training Objectives}\label{train_obj} 
The ultimate goal of the network is to discern whether an observation contains a pulsar or not.
For this goal we use the {\it cross-entropy loss} to train our network which measures the classification performance of our network. This loss measures the performance of the classification and is the loss function most commonly used in classification problems.
Each classifier is trained individually but also the final classification result is added to the loss. 

We add an intermediate loss function that allows the network to find suitable weights for the dedispersion network faster than the classification loss alone would enable. The target function is calculated as follows; the simulated filterbank data is dedispersed and the positions of the peak identified, these locations are convolved with a Gaussian kernel to give the desired signal.The \emph{reconstruction loss} is then calculated as the mean squared loss (MSE) between the output of the dedispersion branch the target signal described above.

When the dedispersion network has multiple output channels we train each channel to be receptive to a part of the whole DM range of the training set. For this we only contain the pulses in the target function of the channel when we want it to be receptive to that particular DM. Otherwise the target function of the channel only contains zeroes.

The loss function that is used to train the network is a combination of the reconstruction loss, the classification loss of the individual classifiers and the classification loss of the combination of the classifiers.
Initially the weight of the reconstruction loss is high but since our final goal is a good classification of the data the weight of the classification loss in the combined loss is increased during training. The details how these weights change during training are described in Section \ref{sec:training}.

%% file: tables/ffa_para.tex
\begin{table}
\begin{tabular}{rrrrr}
Segment & period\_min [s] & period\_max [s] & bins\_min & bins\_max  \\ \hline
1 & 0.03 & 0.12 & 10 & 14 \\
2 & 0.12 & 0.48 & 40 & 44 \\
3 & 0.48 & 1.1 & 160 & 176
\end{tabular}
\caption{Parameters of the FFA calculation. The computed maximum period is a few milliseconds higher than period\_max which leads to a slight overlap between the segments. 
The period range is based on the simulated periods in the training set and the periods of the real pulsars included in the test set. 
The parameters bins\_min and bins\_max are the minimum and the maximum number of phase bins of the folded pulse profile.}
\label{tab:ffa_para}
\end{table}

%% file: models.tex
In this work we experimented with three flavours of the neural network model to tease apart their strengths and weaknesses.
\begin{itemize}
\item The two-channel model is as shown in Figure \ref{fig:architecture} and described in Section \ref{sec:arch}. This model has two output channels after the dedispersing part of the network.

\item The one-channel model is the same as the 2-channel model but only uses one output channel produced by the dedispersing part of the network.

\item The two-step model has the same architecture as the 1-channel model but the dedispersing part and the classification part of the network are trained separately. No gradients from the classification loss are used in the training of the dedispersing part of the network.
\end{itemize}

In the two-channel model the target for the reconstruction loss is constructed in a way where one channel contains the high-DM pulsars and the other channel contains the low-DM pulsars. For the middle 25\% of the DM range of the training data the training target shows peaks in both output channels while for the DM value above and below that range these peaks are only shown in one channel of the training target and the other channel of the training target contains only zeros.

%% file: perf_metrics.tex
The neural network is trained using the mean squared error loss for the intermediate output of the network and the cross entropy loss for the classification output of the network. While these measures are useful to train our network, they are not ideal to judge the performance on real observations.

\paragraph*{Strength of Dedispersed Pulses}
The way the target for the mean squared error loss is created it cannot easily be adapted to real pulsars, since the position of the individual pulses has to be known. In order to investigate how well the network is able to recreate real pulsar pulses we instead use the FFA library \textsc{riptide}. If the network is able to see the pulses of a real pulsar the FFA algorithm will show peaks at the pulse period and its harmonics. The amplitude of the peak indicates how strong the pulsar is in the output of the dedispersion network, compared to other noise. 
To do a meaningful comparison we use the observed DM of the pulsar to create the dedispersed time series of the pulsar observation. We also downsample this time series by a factor of 4 and truncate it to have the same resolution and length as the neural network output. We also apply the FFA to this time series. Comparing the two amplitudes at the pulse period allows us to measure how well our network performs at detecting dispersed signals. In a real survey the DM of the pulsar is not a priori known, which means many DM trials have to be computed. In this comparison we effectively compare our neural network output to best possible such DM trial (i.e. the one that is dedispersed at the actual DM of the pulsar). In other DM trials the S/N of the pulsar will usually be weaker due to smearing.
In Section \ref{sec:ffa_perf} we discuss the performance of the dedispersion network using the FFA.

It should be noted that the various downsampling steps that were applied during the processing may reduce the sensitivity of the FFA output, which is why the computed S/N values may underestimate the capability of a pure FFA pipeline to detect these pulsars.

\paragraph*{Classification}
Since we have a different number of observations with and without pulsars in the test, the cross entropy loss does not necessarily give the best information on how good our network is at detecting pulsars. In order to measure the performance of  a classifier with a single value we use the PyCM library \citep{2018JOSS....3..729H} to calculate the {\it Matthews correlation coefficient} \citep[MCC; originally used in][]{MATTHEWS1975442}
which reaches 1 for a perfect classifier.
The MCC can be calculated from the confusion matrix which contains the values 
TP, TN, FP and FN indicating the true positive, the true negative, the false positive and the false negative classification results.
The MCC can be calculated by \citep[][]{baldi2000assessing}:
\begin{equation}
\label{eq:mcc}
\text{MCC} = \frac{ \mathit{TP} \times \mathit{TN} - \mathit{FP} \times \mathit{FN} } {\sqrt{ (\mathit{TP} + \mathit{FP}) ( \mathit{TP} + \mathit{FN} ) ( \mathit{TN} + \mathit{FP} ) ( \mathit{TN} + \mathit{FN} ) } }
\end{equation}

To improve our classification performance we split the observation in five largely overlapping segments. When two of these classify the observation as containing a pulsar, the whole observation is classified as containing a pulsar. The overlap between the segments depends on the input size of the neural network.

To describe the prediction of the models on a particular pulsar we compute the softmax function of the combined prediction. We call the output value for the pulsar class {\it pulsar prediction}. This value reaches 1 when the network is very sure that the observations contains a  pulsar and 0 if it does not see a pulsar.

In sections \ref{sec:perf_fake} and \ref{sec:perf_real} we discuss the classification performance of the neural networks on both simulated and real pulsar observations.

%% file: training.tex
 \label{sec:training}

We use a novel technique to pre-calibrate the neural network to be attuned to the detection of radio pulses. Because radio pulsars are very faint, we start training the network with disproportionately strong signals, and gradually diminish their strength as the training progresses so as to enable the network to reconstruct the pulses. 
The inputs to training are real observations from a survey  containing noise and RFI to which we add a simulated pulsar as shown in Equation~\ref{eq:I_combine} (see Section~\ref{sec:simulations}).
For training and validations we use real observations which to our knowledge do not contain a pulsar and inject a simulated pulsar in 50\% of the samples.
The noise level $\mathcal{N}$ is sampled from a uniform distribution during the training step.
During validation the lower end of this distribution is used as the noise level. If the network achieves a higher MCC than 0.85 on the validation set in three subsequent epochs, the lower level of the noise distribution is increased. When this happens the optimizer of the network is reset. This scheme allows the network to quickly converge to useful weights using strong signals and later on adapt to weak signals.

Our neural networks are trained in three steps:
\begin{enumerate}[i]
\item the input has 100,000 time steps and the noise level $N$ is increased from 0.9 to 15 during the training. Both the classification (MSE) and the reconstruction (cross-entropy)  losses are used for training. For the total loss calculation, the classification loss is weighted by 0.01 in order for the network to focus on getting the reconstruction right. For very strong pulsars a short input length suffices to make the right classification. The small input length allows us to quickly iterate and find useful weights for the convolutional neural net. The classification is done using the FFT and STFT classifier modules. 
\item The input length is increased to 400,000. The dedispersing part of the network is initially retained from step (i), but the classifiers are re-trained from scratch because their architecture is a function of the input size, which has now changed. Again we only use the FFT and STFT classifiers. During this training step the noise level is gradually increased from 15 to 35 as the network improves its performance at a given noise level. In this step the weight of the classification loss is 0.1 and the weight of the reconstruction loss is again one. During the first two epochs \footnote{ An epoch is defined as training the neural network on the entire training set once.} of training the weights of the dedispersing part of the network are frozen since the weights of the classifiers are initialised randomly and could possibly disrupt training. 
\item Finally, the FFA classifier is added to the network. Since in the current implementation we do not propagate gradients through the FFA classifier we freeze the weights of the dedispersing network and only train the classifiers and the weights for the individual classifications that are combined into a final classification. During this step the noise level was held fixed at the highest noise level reached during the previous step \footnote{Not all networks reach a noise level of 35 during the second training step since the performance on the validation set is in some cases not good enough.}.
\end{enumerate}

The training of the neural networks was done on the Bielefeld GPU cluster which uses NVIDIA Tesla V100 with 32 GB memory. The training of one neural network was performed on a single GPU where we used a batch size of 15 in the second and third training step meaning we apply our model to 15 observations at once during training. 
When using the two-channel model the first training was finished in less than 3 hours, the second training step was finished in less than 13 hours and the third training step took about 50 hours. Taking the data from the GPU to the CPU and applying the FFA algorithm is the biggest time cost. Getting a single prediction for all 105 files in the test set when using the FFA classifier takes roughly 105 seconds and without the FFA classifier takes about 29 seconds.


Since the total observations only consist of 418,751 time steps, in the second training step the neural network uses the majority of the pulsar observations for the classification. When classifying real pulsar observations the neural network looks at five overlapping segments of the real observations. In this case the overlap is very large and the neural network mostly looks at the same data in the different segments.

For the training of the network we use pulsar survey observations which do not contain pulsars and simulated pulsars contained in the training+validation set. 
20\% of the training+validation set is used as the validation set in each training run. 
To reduce the effect of noisy training, a five-fold cross-validation test is performed where different non-overlapping sets of the training+validation set are used as validation sets.
In Table \ref{tab:set_size} the sizes of the data sets used for training and validation are outlined. The data sets are further described in Section \ref{sec:data}.

%% file: tables/set_size.tex
\begin{table}
\begin{tabular}{lcc|cc}
                & \multicolumn{2}{c|}{Training+validation set} & \multicolumn{2}{c}{Test set} \\ \hline
                & Observations     & Simulations     & Pulsars     & No Pulsars     \\
Number of files & 529              & 8,620           & 24          & 81            
\end{tabular}
\caption{The sizes of the data sets used to train and test the performance of the neural network. The training+validation set consists of simulations and observations that do not contain pulsars while the test set only consists of real observations which do or do not contain pulsars. For each training run 80\% of the training/validation set are used for training and the rest is used for validation. We perform a five-fold cross-validation test with non-overlapping validation sets.}
\label{tab:set_size}
\end{table}

%% file: tables/psr_table_pred.tex
\begin{landscape}
\clearpage

\begin{table}    

\begin{tabular}{cp{1.8cm}>{\raggedleft\arraybackslash}p{1.8cm}>{\raggedleft\arraybackslash}p{1.6cm}>{\raggedleft\arraybackslash}p{1.4cm}>{\raggedleft\arraybackslash}p{2cm}>{\raggedleft\arraybackslash}p{2cm}>{\raggedleft\arraybackslash}p{2cm}>{\raggedleft\arraybackslash}p{1.8cm}>{\raggedleft\arraybackslash}p{1.8cm}>{\raggedleft\arraybackslash}p{1.8cm}}
    &     \centering PSR \newline JName &  \centering DM \newline (pc cm$^{-3}$) & \centering Period \newline (ms) &  \centering S/N$_{FFA}$ \newline Optimal \newline Dedispersion &  \centering S/N$_{FFA}$ \newline 2-channel \newline model & \centering S/N$_{FFA}$ \newline 1-channel \newline model & \centering S/N$_{FFA}$ \newline 2-step \newline model &  \centering Prediction \newline 2-channel model & \centering Prediction \newline 1-channel model & \centering Prediction \newline 2-step \newline model \cr\\ \hline
  1 &  J1902+0615 &                                  502 &                            673 &                                        110.9 &                                  133.0 &                                  139.2 &                                  152.7 &                                  1.00 &                                  1.00 &                                  1.00 \\
  2 &  J1957+2831 &                                  138 &                            307 &                                         74.8 &                                  135.2 &                                  126.2 &                                  146.6 &                                  1.00 &                                  1.00 &                                  1.00 \\
  3 &  J1850+0026 &                                  201 &                           1081 &                                         53.3 &                                   92.5 &                                   83.0 &                                   98.1 &                                  1.00 &                                  1.00 &                                  1.00 \\
  4 &  J1948+2551 &                                  289 &                            196 &                                         53.0 &                                   67.3 &                                   65.6 &                                   82.0 &                                  1.00 &                                  1.00 &                                  1.00 \\
  5 &  J1948+2333 &                                  198 &                            528 &                                         35.3 &                                   43.1 &                                   36.3 &                                   46.2 &                                  1.00 &                                  1.00 &                                  1.00 \\
  6 &  J2005+3547 &                                  401 &                            615 &                                         25.8 &                                   34.7 &                                   31.5 &                                   52.3 &                                  1.00 &                                  1.00 &                                  1.00 \\
  7 &  J1856+0245 &                                  623 &                             80 &                                         21.4 &                                   16.0 &                                    5.1 &                                    6.5 &                                  1.00 &                                  0.10 &                                  0.73 \\
  8 &  J2005+3552 &                                  455 &                            307 &                                         15.7 &                                   20.0 &                                   15.1 &                                   15.1 &                                  1.00 &                                  1.00 &                                  0.97 \\
  9 &  J1855+0307 &                                  402 &                            845 &                                         15.5 &                                   19.1 &                                   16.7 &                                    5.6 &                                  1.00 &                                  1.00 &                                  0.12 \\
 10 &  J2013+3058 &                                  148 &                            276 &                                         14.0 &                                   15.7 &                                   13.5 &                                    5.3 &                                  1.00 &                                  1.00 &                                  0.07 \\
 11 &  J1947+1957 &                                  185 &                            157 &                                         13.4 &                                   13.8 &                                   12.2 &                                    5.3 &                                  1.00 &                                  1.00 &                                  0.08 \\
 12 &  J1902+0615 &                                  502 &                            673 &                                         11.4 &                                   12.7 &                                    7.7 &                                    1.1 &                                  1.00 &                                  0.67 &                                  0.11 \\
 13 &  J1850+0423 &                                  265 &                            290 &                                         11.4 &                                   16.1 &                                   14.7 &                                    5.8 &                                  1.00 &                                  1.00 &                                  0.06 \\
 14 &  J1934+2352 &                                  355 &                            178 &                                         11.2 &                                    7.8 &                                    8.1 &                                    5.1 &                                  0.85 &                                  0.95 &                                  0.09 \\
 15 &  J1916+1225 &                                  265 &                            227 &                                         10.1 &                                    8.8 &                                    8.4 &                                    4.9 &                                  1.00 &                                  0.98 &                                  0.11 \\
 16 &  J1906+0641 &                                  472 &                            267 &                                          9.8 &                                   11.6 &                                    9.4 &                                    5.5 &                                  1.00 &                                  1.00 &                                  0.08 \\
 17 &  J1859+0603 &                                  378 &                            508 &                                          9.8 &                                   12.4 &                                   13.0 &                                    7.1 &                                  1.00 &                                  1.00 &                                  0.14 \\
 18 &  J2007+3120 &                                  191 &                            608 &                                          9.4 &                                   12.7 &                                    9.3 &                                    4.4 &                                  1.00 &                                  1.00 &                                  0.08 \\
 19 &  J1946+2535 &                                  248 &                            515 &                                          9.3 &                                    9.4 &                                    9.0 &                                    4.4 &                                  0.95 &                                  0.99 &                                  0.08 \\
 20 &  J1848+0351 &                                  336 &                            191 &                                          9.1 &                                    8.0 &                                    8.3 &                                    5.1 &                                  0.90 &                                  0.97 &                                  0.11 \\
 21 &  J1906+0641 &                                  472 &                            267 &                                          8.7 &                                    9.6 &                                    6.5 &                                    4.3 &                                  1.00 &                                  0.34 &                                  0.15 \\
 22 &  J1858+0346 &                                  386 &                            256 &                                          8.0 &                                   10.9 &                                   10.8 &                                    5.9 &                                  1.00 &                                  1.00 &                                  0.10 \\
 23 &  J1908+0909 &                                  467 &                            336 &                                          5.4 &                                    4.5 &                                    4.1 &                                    3.4 &                                  0.08 &                                  0.08 &                                  0.07 \\
 24 &  J1851+0418 &                                  115 &                            284 &                                          5.3 &                                    6.1 &                                    5.2 &                                    4.7 &                                  0.04 &                                  0.06 &                                  0.09 \\
\end{tabular}
\caption{Pulsars in the test set sorted by their S/N in the FFA of the optimally dedispersed output and the predictions of the neural networks. 
When the network has multiple output channels the stronger signal was used. For the calculation of the S/N$_{FFA}$ the first 400 000 time steps of the observation were given to the neural networks. The optimally dedispersed output was created after removing the mean of each time step and was downsampled by a factor of four after dedispersing. Only the first 100 000 time steps are used in the FFA calculation in order to have the same input size to the FFA as in the neural network case.  Since for each model five networks where trained the median is shown for the S/N$_{FFA}$. The resulting FFA peak values are discussed in Section \ref{sec:ffa_perf}.
The pulsar prediction value is the value for the pulsar class after calculating the softmax of the output of the combined classifier and is discussed in Section \ref{sec:perf_real}. For the prediction five overlapping segments of each observation were given to the networks and for each network the median prediction was calculated. Afterwards the median of the median prediction of the individual networks was calculated for each model.}
\label{tab:psr}
\end{table}
\end{landscape}
\newpage

%% file: observations.tex
Our neural network model is tested on publicly available data of the PALFA survey \citep{2015ApJ...812...81L} that have been recorded with the Wideband Arecibo Pulsar Processor (WAPP) \citep{2000ASPC..202..275D}. These contain 256 frequency channels, spanning a bandwidth of 100 MHz and a time resolution of 64 $\mu s$. The observations span 268 seconds each. Before processing the data, we downsampled it to 16 channels and a time resolution of 640 $\mu s$ resulting in 400,000 time steps. The lowest and highest frequency channel of the downsampled signal are discarded because of poor sensitivity.
The WAPP was only used in the early stages of the PALFA survey and has a smaller dynamic range than later processing backends.
We used data from the early stages of this survey due to its public availability and the small size of individual observations.
Using data from a survey with a low data volume but with high sensitivity to real pulsars
due to the large collecting area of the Arecibo telescope
allowed us to test a variety of different approaches how to set up our model.
Using a further downsampled version of the data allows faster prototyping of the model and does not hugely influence the sensitivity to non-millisecond pulsars. 

As noise samples for the training of the network we used 529 beams that were observed between MJDs 53830 and 53832. To test for pulsars in these observation we checked if there are known pulsars withing a beam width of the pointing. We folded the observations with the period and DM of the known pulsar and checked by eye if a pulsar is discernible in the data. This resulted in the redetection of nine pulsars. Additionally we use 15 pulsar discovery observations from other dates that were observed with the same system set-up.

This results in 24 pointings containing pulsars which we use to test the performance of the neural network on real data. The specifics of the pulsars are listed in Table \ref{tab:psr}. Our test set also includes the other 81 beams of the discovery observations, which do not contain known pulsars.

%% file: simulations.tex
 \label{sec:simulations}
Training the neural network is done by giving it segments of filterbank files which either contain a simulated pulsar embedded in noise or only noise. The task of the network is differentiating these two cases.

Training data have been created with the \textsc{ SKA-TestVectorGenerationPipeline}\footnote{\hypertarget{https://github.com/scienceguyrob/SKA-TestVectorGenerationPipeline}{https://github.com/scienceguyrob/SKA-TestVectorGenerationPipeline}} 
package which utilises the program \textsc{inject\_pulsar} contained in \textsc{sigproc}\footnote{\hypertarget{https://github.com/SixByNine/sigproc}{https://github.com/SixByNine/sigproc}, 
originally described in \citet{2011ascl.soft07016L}}. The pipeline uses profiles of the EPN database \footnote{\hypertarget{http://www.epta.eu.org/epndb/}{http://www.epta.eu.org/epndb/}} to create simulated pulses. These simulated pulses are injected into empty filterbank files.

The pulsar periods of the simulated data set are uniformly distributed between 30 ms and 650 ms while DM values are uniformly distributed between 80 and 700. Originally 10,000 simulated files were created with a chosen target S/N parameter of 70. Using this simulation pipeline resulted in a number of simulated pulsars having much stronger individual pulses than others. Since very strong pulses are trivial to classify we did not use any simulated files where the maximum values exceeded 20. This results in a set of 8,620 simulated pulsars.

During training the simulated pulses are added to real observations after dividing them by the noise level $\mathcal{N}$ which at the start of the training is 0.9 and increases to 35:

\begin{equation}
\label{eq:I_combine}
I_{combined} = I_{noise} + \frac{I_{simulated}}{\mathcal{N}}
\end{equation}

The simulated files contain intensities $I_{simulated}$ between 0 and 20 while a typical noise observation contains intensities $I_{noise}$ between 55 and 75. Since we increase $\mathcal{N}$ to 30 at the end of the training the intensity of the pulse in each pixel is smaller than the quantization of the initial filterbank file.

Inevitably this method of creating fake pulsars does increase the mean value of the observation slightly and also creates floating point values that were not included in the original 8 bit filterbank file. This poses the threat that the networks learn to identify the files containing simulated pulsars by these non-integer values. This cannot be completely ruled out but we have seen that the network is also able to generalize to real pulsar observations. The increase in mean value is counteracted by normalising the data as described in Section \ref{sec:preprocess}.

%% file: ffa_performance.tex
\label{sec:ffa_perf}

One way to test if the transformations that happen in the neural network are physically meaningful is to look at an intermediate output, for example, the one that we trained to become the dedispersed signal, and apply traditional search techniques from there onward to see if a pulsar is contained in that output. In this work we use the FFA to investigate the quality of the output of the dedispersion network.

The individual data points in the FFA measure the S/N of the input data when these are folded at a certain period. Although the periodic pulsar signals create peaks in the FFA, unfortunately so does the RFI. We compare the performance of our neural network against dedispersing the (already downsampled) test set of pulsar observation at the known pulsar DM and afterwards downsampling the data by factor of 4 to create a time series with the same length as the output of the dedispersion network.
When dedispersing the data at the observed pulsar DM we henceforth call this the {\it optimal dedispersion}. 

The presence of RFI creates structures in the FFA which make the pulses resulting from peaks less easily discernible. To mitigate the effect of the RFI we can subtract the mean of each time point across all frequencies from the dispersed signal. Because the RFI has no DM (i.e., 0-DM) it appears at the same location in time across all frequencies and subtracting the mean removes a large fraction of the RFI \citep{2009MNRAS.395..410E}.

The FFA library \textsc{riptide} uses a local significance threshold for the prediction of peaks. The threshold is given by a polynomic fit through $s=m+k\sigma$ where $m$ is the the local median, $\sigma$ is the local standard deviation and $k$ is a dimensionless constant defining the significance level. The details of the calculation can be found in Section 5.5 of \citet{2020MNRAS.497.4654M}. We use $k=6.5$ and $smin=6.6$ which is the minimum value of the threshold.
We use the calculated threshold to correct the S/N amplitudes for a change in the noise floor and level of the FFA. As can be seen in Figure \ref{fig:ffa_rfi_17} the noise level of the FFA differs between neural network dedispersion, optimal dedispersion and optimal dedispersion with zero-DM filtering. 
We subtract the local significance threshold from the measured amplitude and add the minimum value of 6.5 to it afterwards.
This will reduce the amplitude of the FFA when the quality of the FFA suffers from the presence of RFI and give a better representation of the significance of the signal than just the amplitude alone.

\begin{figure}
 \includegraphics[width=\columnwidth]{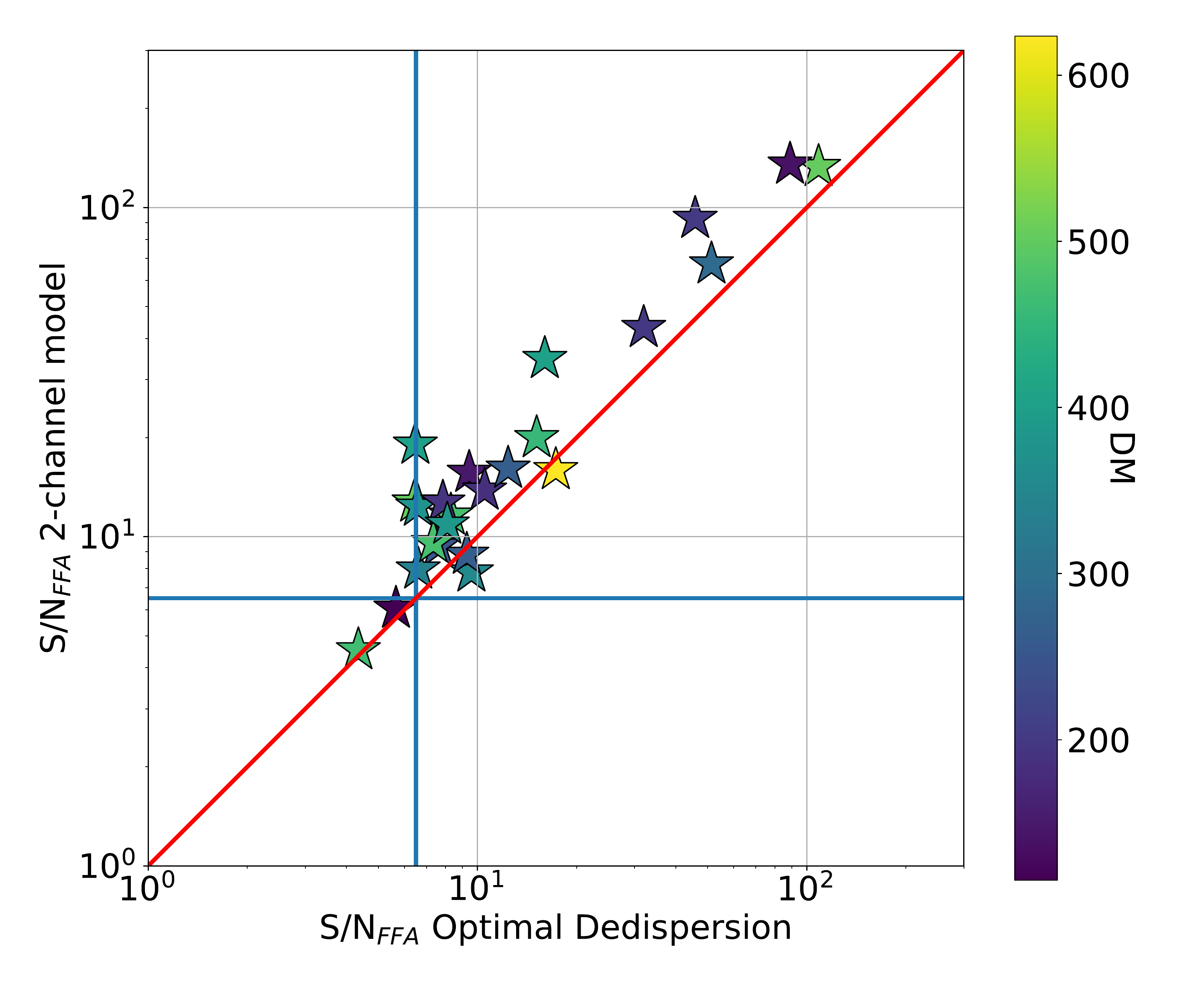}
 \caption{Comparison of the S/N in the FFA when using the two-channel model or when dedispersing the test pulsar at the pulsar DM. The horizontal and vertical bold lines show thresholds of 6.5. We see that the two-channel model is competitive with optimal dedispersion, it is even slightly better, presumably due to RFI mitigation in the neural network.}
 \label{fig:scatter_m0_dm}
\end{figure}

\begin{figure}
 \includegraphics[width=\columnwidth]{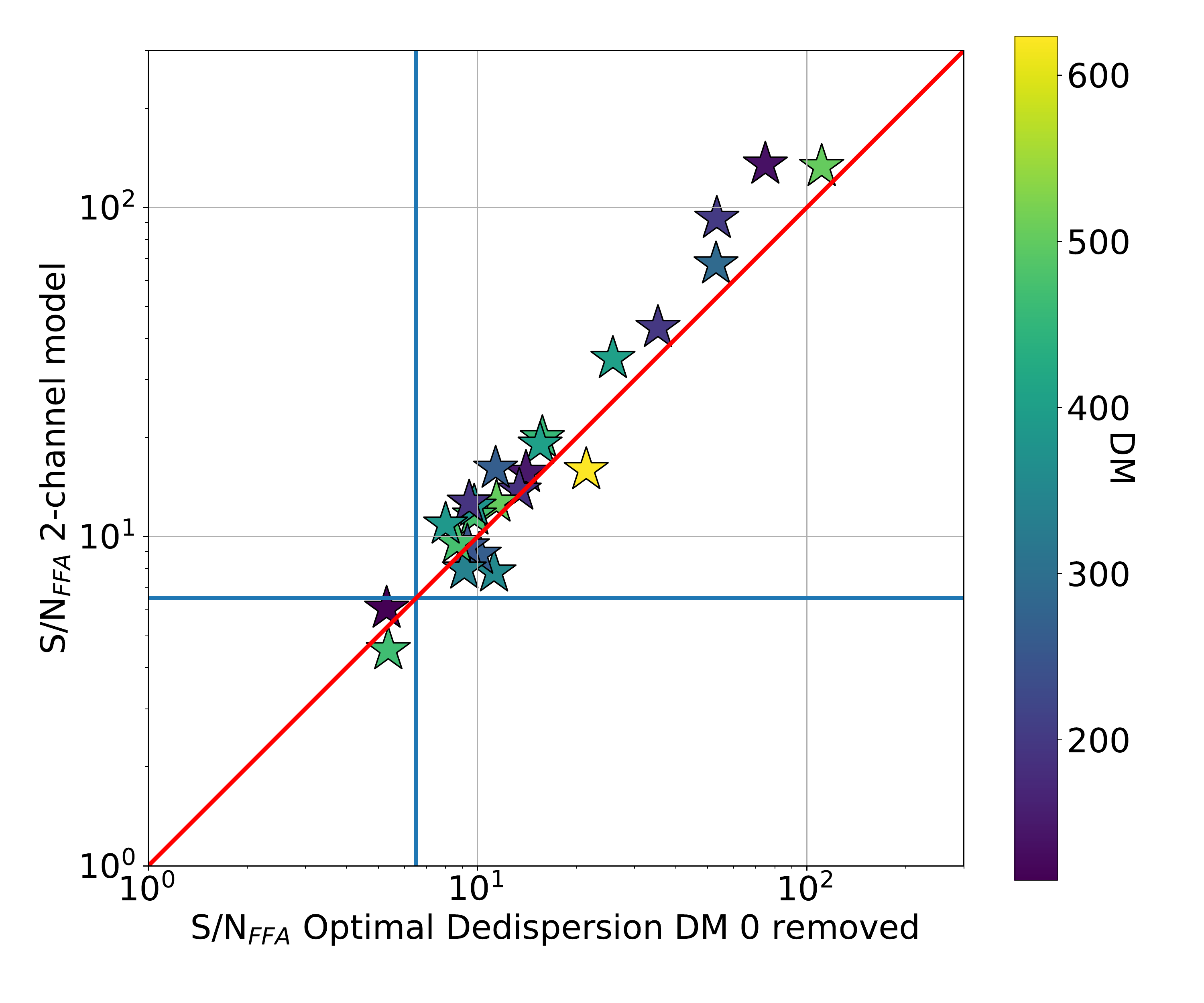}
 \caption{Comparison of the S/N in the FFA when using the two-channel model and when dedispersing the test pulsar at the pulsar DM after removal of undispersed signals. The bold lines are as in Figure \ref{fig:scatter_m0_dm}. Comparing with Figure \ref{fig:scatter_m0_dm} shows that the performance of the neural network and optimally dedispersing is more comparable when using a form of RFI suppression when optimally dedispersing.}
 \label{fig:scatter_m0_dmrm0}
\end{figure}

Figure \ref{fig:scatter_m0_dm} shows a comparison of the amplitudes of the pulsar signal after doing the described threshold removal when comparing the two-channel model to the dedispersed output without RFI removal. Figure \ref{fig:scatter_m0_dmrm0} shows the same comparison with the described RFI removal technique. We see that the dedispersed time series where broadband RFI has been suppressed provides a stronger base line to compare our network against. When comparing our network with the dedispersed output without RFI removal we observe that our network is able to suppress RFI to an extent which is more closely described in Section \ref{sec:rfi}. We observe that two pulsars are below the default threshold of 6.5 and are hard to detect in these observations even when the DM is known.

We see that using our neural network is competitive with dedispersing the pulsar at the known DM. For the majority of pulsars we observe a slightly stronger signal in the neural network output. This proves the capability of the network to be sensitive to dispersed pulses and to effectively ignore RFI.

The strength of the pulsar peaks in the FFA of the intermediate output and the dedispersed time series is shown in Table \ref{tab:psr}. 

\begin{figure}
 \includegraphics[width=\columnwidth]{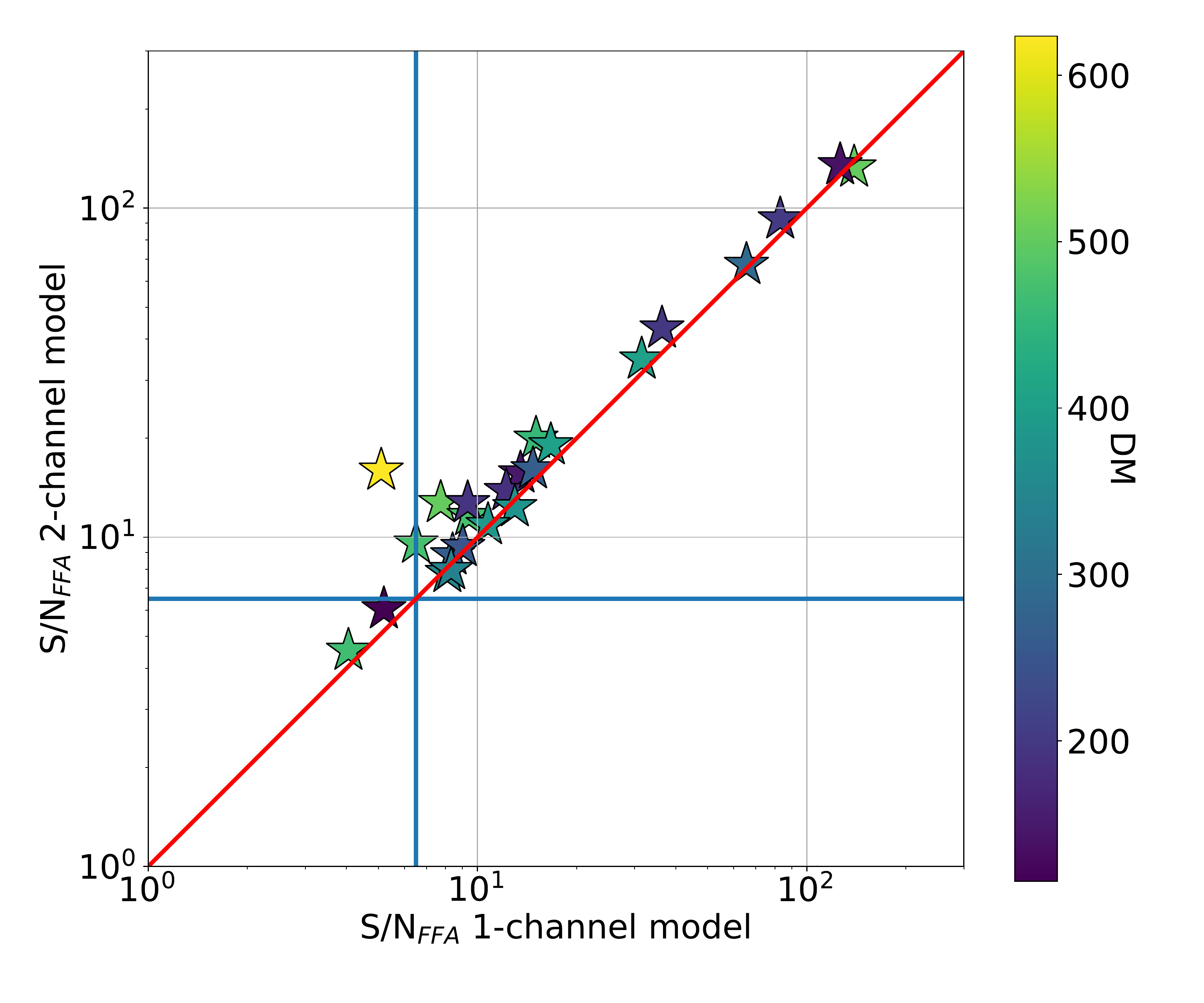}
 \caption{Comparison of the pulsar S/N$_{FFA}$ between the two-channel model and the one-channel model. The bold lines are as in Figure \ref{fig:scatter_m0_dm}. For the two-channel model the maximum value of the two resulting FFAs is given. The one-channel model shows a weak output for the pulsar with the highest DM in the test data set.}
 \label{fig:ffa_model12}
\end{figure}

Figure \ref{fig:ffa_model12} shows a comparison between the two-channel model and the one-channel model. The model using two output channels shows higher maximum values in the FFA. Especially the pulsar with the highest DM is weaker when using only one output channel. This shows that the network is able to compress a range of signals coming from various DMs into a single output channel, but giving it the opportunity to separate low and high DM signals in different channels may result in a more useful output.

\begin{figure}
 \includegraphics[width=\columnwidth]{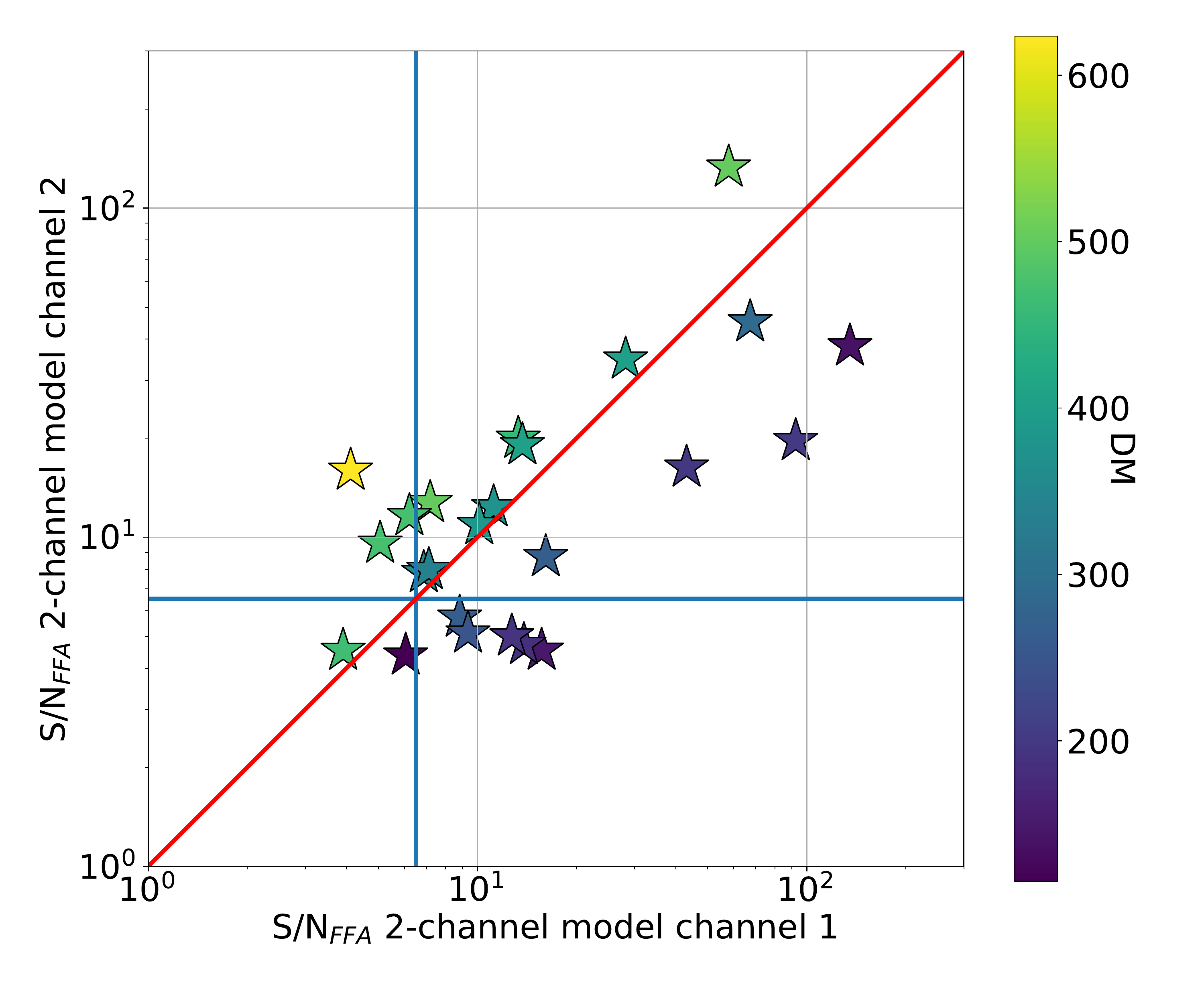}
 \caption{Comparison of the pulsar S/N in the FFA between the two channels of the two-channel model. The bold lines are as in Figure \ref{fig:scatter_m0_dm}. Many pulsars are clearly visible in both channels but the low-DM pulsars are stronger in one channel and the high-DM pulsars are stronger in the other channel.}
 \label{fig:ffa_model1_chan}
\end{figure}

In Figure \ref{fig:ffa_model1_chan} the difference in strength of the pulsars in the two output channels of the two-channel model is shown. Low-DM pulsars are stronger in the first channel while high-DM pulsars are stronger in the second channel. Pulsars at a range of intermediate DMs are detectable in both channels.

\begin{figure}
 \includegraphics[width=\columnwidth]{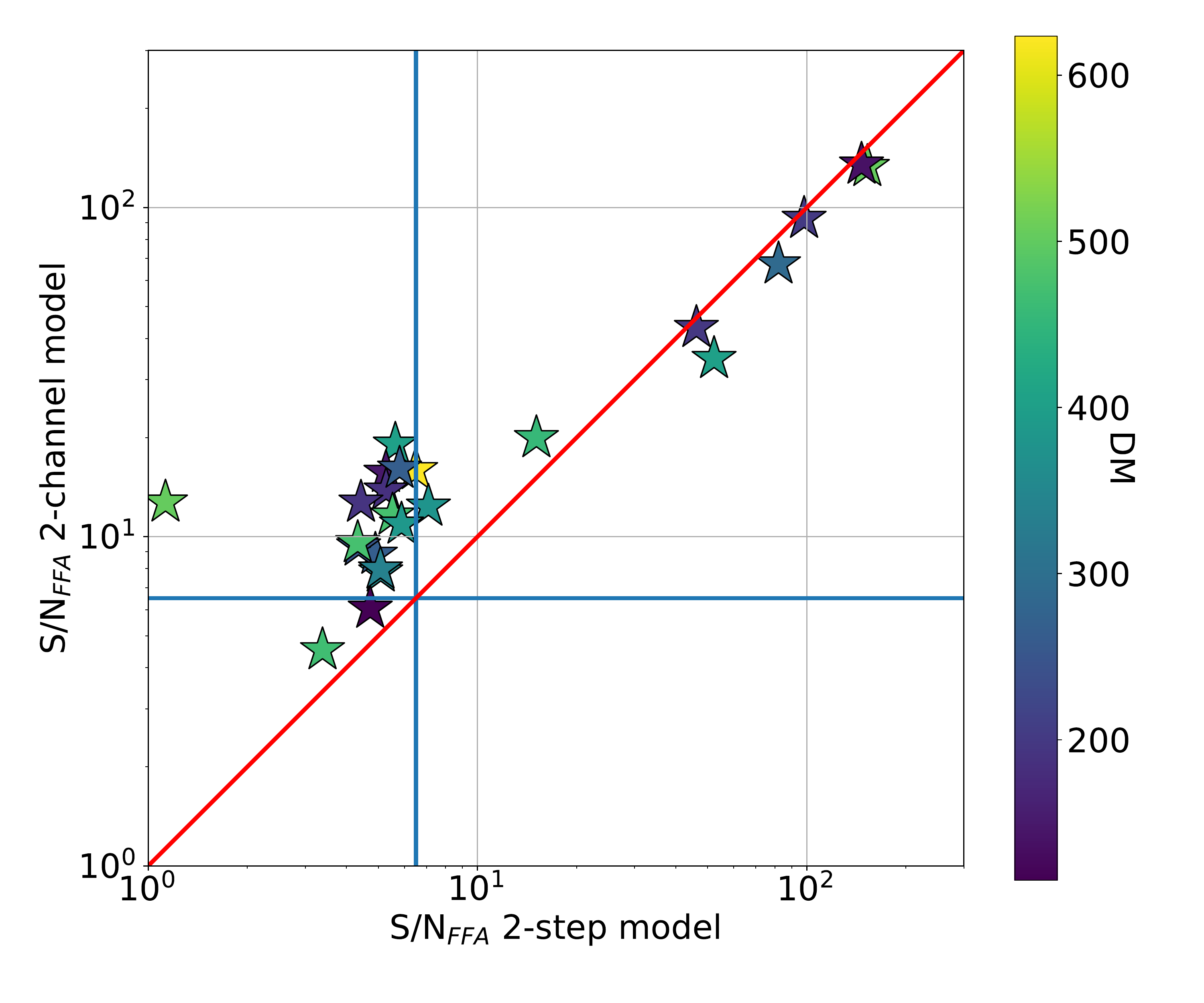}
 \caption{Comparison of the pulsar S/N in the FFA between the two-channel model and the two-step model. Especially for weak pulsars adding the classification loss allows the network to reach a higher S/N in the FFA.}
 \label{fig:ffa_model13}
\end{figure}

Figure \ref{fig:ffa_model13} shows the difference in performance between the two-channel model and the two-step model. Especially for weak pulsars the two-channel model provides a stronger output even though the architecture that creates the intermediate output is the same. This shows that using a classification loss helps creating a more useful intermediate output than solely using the reconstruction loss, at least for relatively faint signals.

%% file: class_performance_fake.tex
\label{sec:perf_fake}

During training the neural network is presented with various combinations of survey observations with simulated pulsars that have been scaled down with an increasing noise level. At the final noise level of 35 during the training the network is no longer capable of optimally discovering all pulsars in the training and validation sets. At this noise level the amplitude of the scaled-down simulated signal is smaller than the quantization precision of the survey observations. 

\begin{figure}
 \includegraphics[width=\columnwidth]{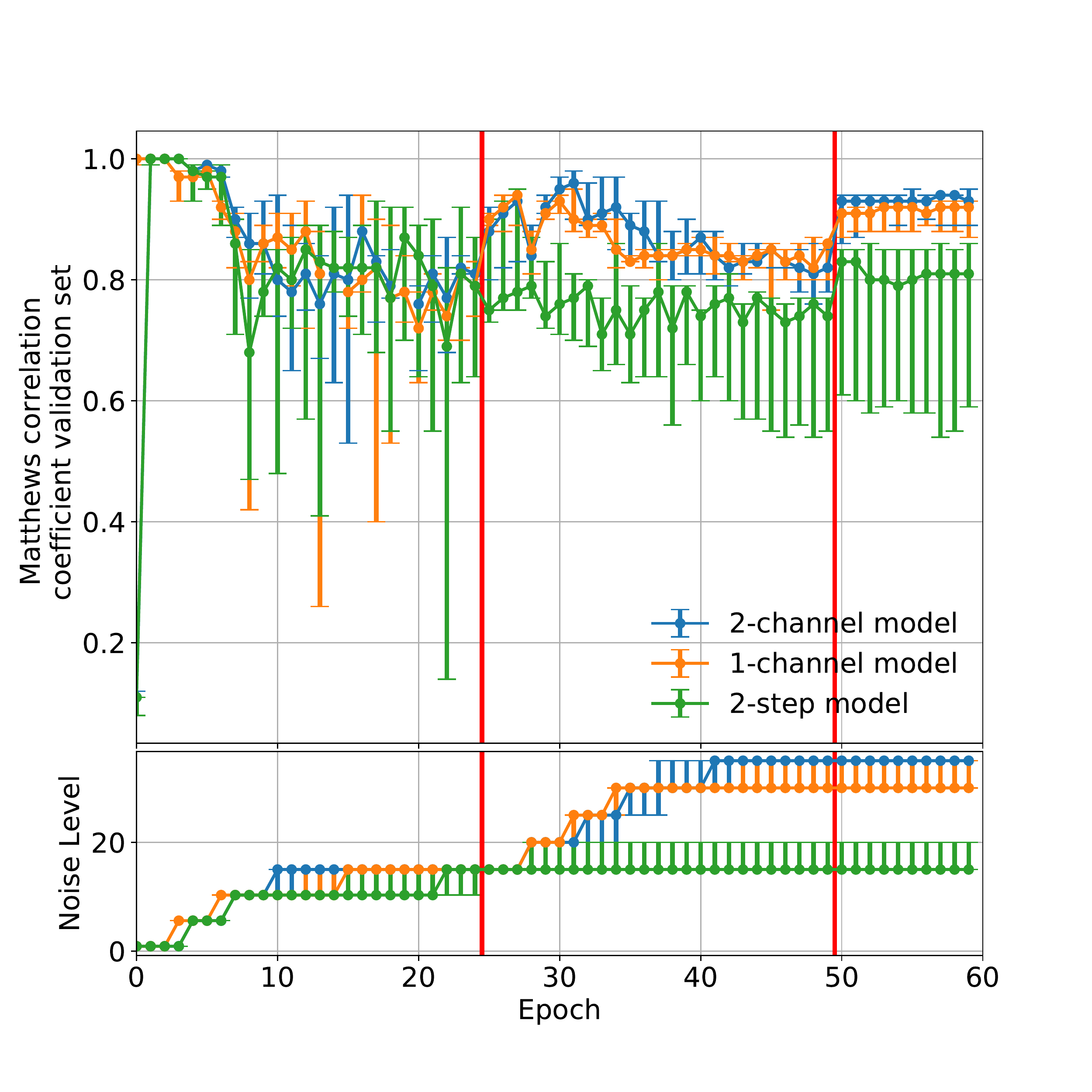}
 \caption{MCC of the validation set and the noise level in the validation data during training of the three models. The plotted values are the values for the median, the minimum and the maximum of the five cross validation runs. The bold vertical lines show the transitions between the three training steps as described in Section \ref{sec:training}. The two-channel model performs slightly better than the one-channel model which can be seen in the more reliably increased noise level after training step two while the two-step model is clearly outperformed by the different training method of the other two models.}
 \label{fig:mcc_dev_val}
\end{figure}

In Figure \ref{fig:mcc_dev_val} the development of the average MCC of the validation set is shown during training. Also shown is the evolution of the noise level of the validation set which increases for each network when it had a MCC of above 0.85 on the validation set for three epochs. Since the noise level of the validation set changes during training one does not see a steady increase in performance as is usual in training neural networks.

The transitions to the second and third training steps (see Section \ref{sec:training}) show clear increases in performance since they introduce a higher input length and a more sensitive FFA classifier.

The comparison between the three models shows that the models using end-to-end training clearly outperform the two-step model. The two-channel model slightly outperforms the one-channel model.

%% file: class_performance_real.tex
\label{sec:perf_real}

A good performance on real pulsar observations is of ultimate importance for the neural network model presented in this work. In this section we investigate how well the neural network is able to adapt from simulated pulsar signals to real pulsar signals.

\begin{figure}
 \includegraphics[width=\columnwidth]{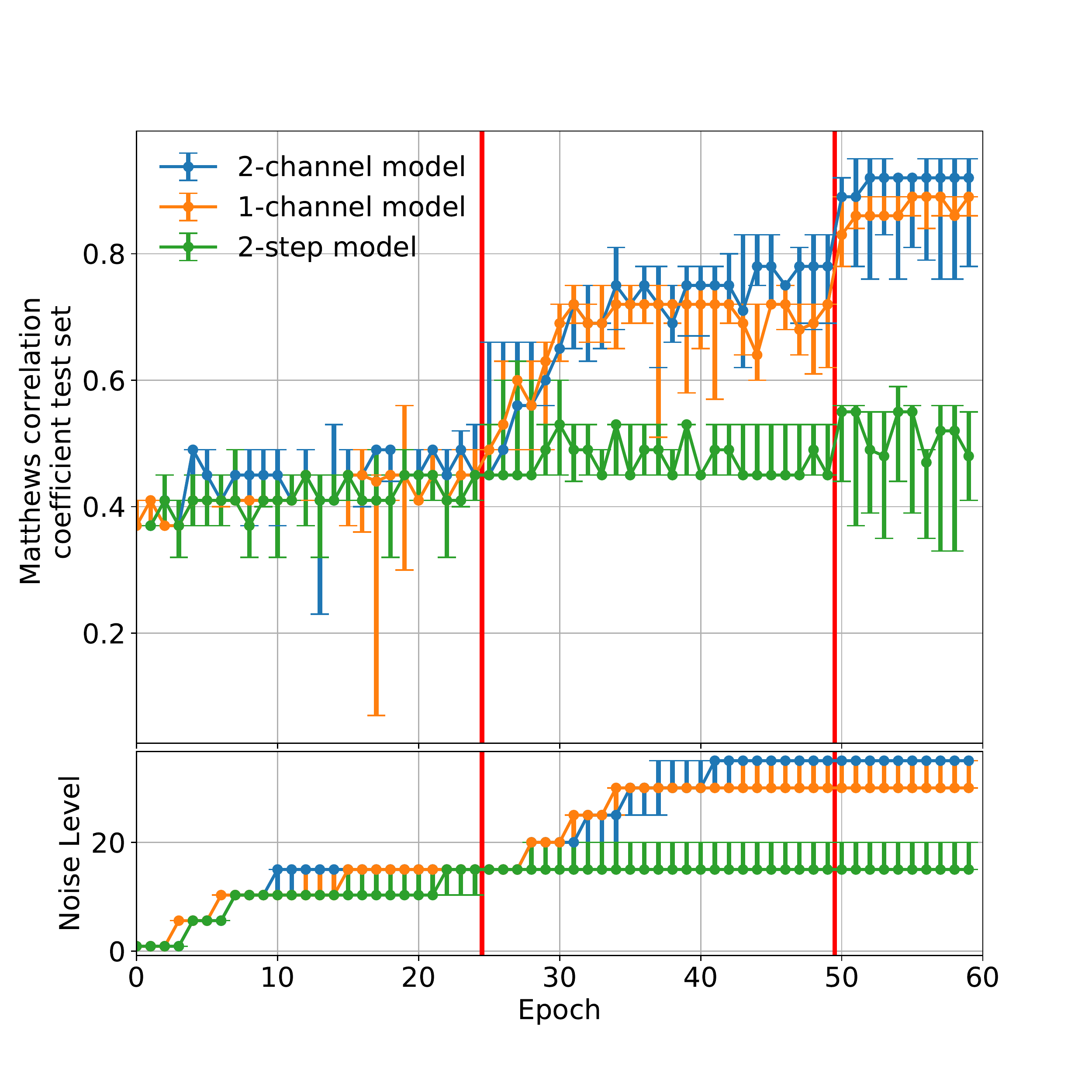}
 \caption{MCC of the test set and the noise level in the validation data during training of the three models. The structure of the plot is the same as in Figure \ref{fig:mcc_dev_val} but with the values for the test set instead. The two-channels model performs slightly better than the one-channel model while the two-step model is clearly outperformed by the different training method of the other two models. Since the number of pulsars is relatively low, the MCC only has a few different values.}
 \label{fig:mcc_dev}
\end{figure}

In Figure \ref{fig:mcc_dev} the development of the MCC of the test data during training is shown for the three models. This graph shows that the networks improve their performance on real data during training. Similar to the performance on simulated pulsars the network profits from the increase in input length and the introduction of the FFA classifier. The mean performance of the two-channel model is better than that of the one-channel model but has a higher variance between the trained models. The two-step model once again is outperformed by the other two models. 

\begin{figure}
 \includegraphics[width=\columnwidth]{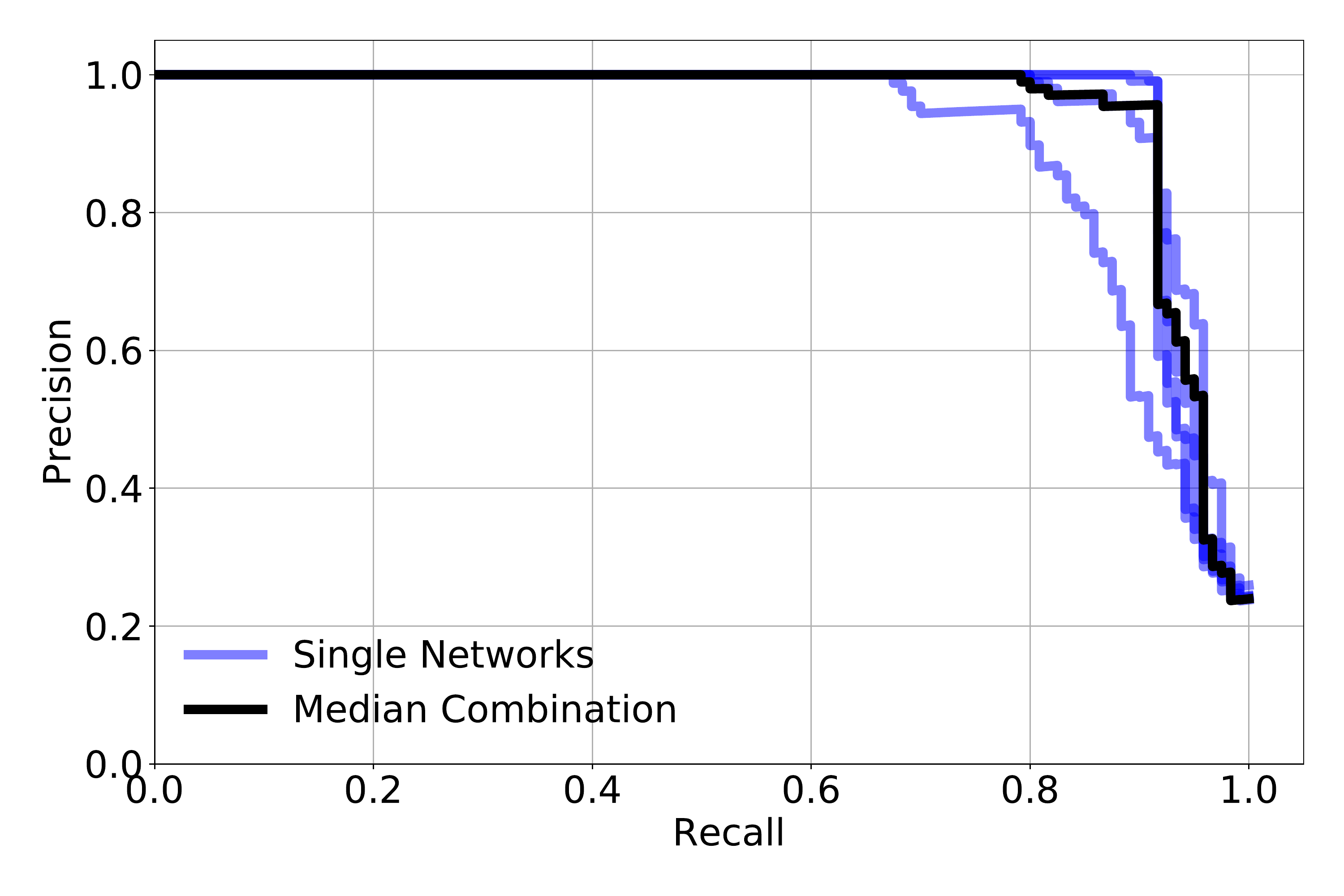}
 \caption{Precision-recall curves for the two-channel model after training. In addition to the result of the individual cross validation runs a combined prediction using the median of the individual predictions is also given.} 
 \label{fig:prec_recall_singles}
\end{figure}

The performance of the individual models which were trained in the cross-validation test of the two-channel model at the end of training are shown in the precision-recall curve in Figure \ref{fig:prec_recall_singles}. While four models perform similarly well, one model performs significantly worse which shows the need to train multiple models to make meaningful comparisons between different architectures.
In the final step of training, four of the neural networks in the cross-validation test of the two-channel model were able to identify 22 out of 24 pulsars while the network that shows the subpar performance in Figure \ref{fig:prec_recall_singles} was only able to detect 20. This network also showed the worst performance on its validation set. In the one-channel model three of the networks were ale to detect 21 pulsars and two only detected 20.

\begin{figure}
 \includegraphics[width=\columnwidth]{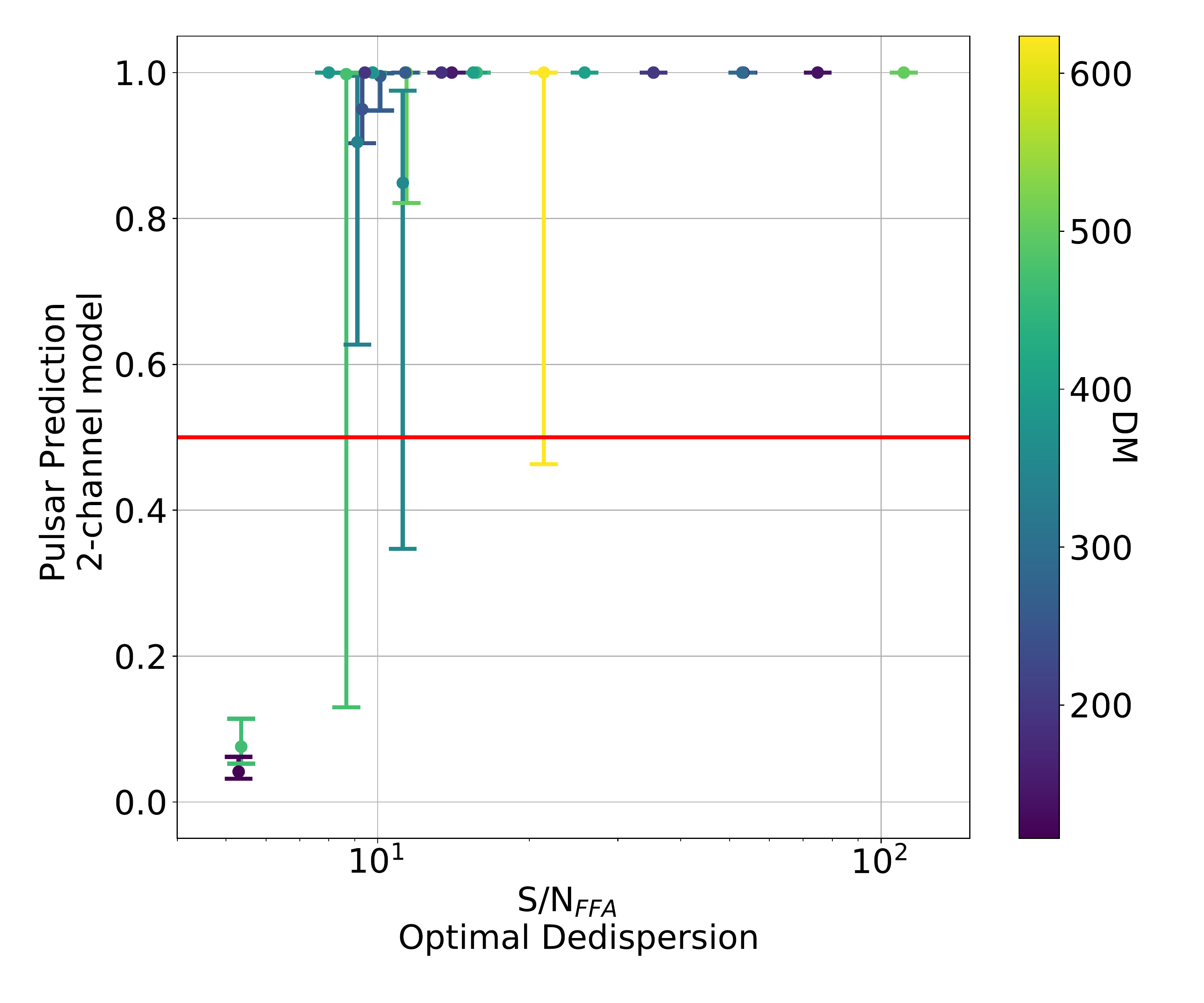}
 \caption{Comparison of the S/N$_{FFA}$ of the optimally dedispersed time series and the Pulsar Prediction of the two-channel model. Five overlapping segments were used to create the median prediction for each individual network. The individual points represent the median of the five trained networks while the error bars indicate the maximum and the minimum of the five median predictions. The minimum prediction is heavily influenced by one particular network which also showed worse performance on its validation set than the other networks. Only the weakest pulsars which have a S/N$_{FFA}$ of 5.3 and 5.4 are not correctly classified by the any of the networks.}
 \label{fig:model1_class}
\end{figure}

\begin{figure}
 \includegraphics[width=\columnwidth]{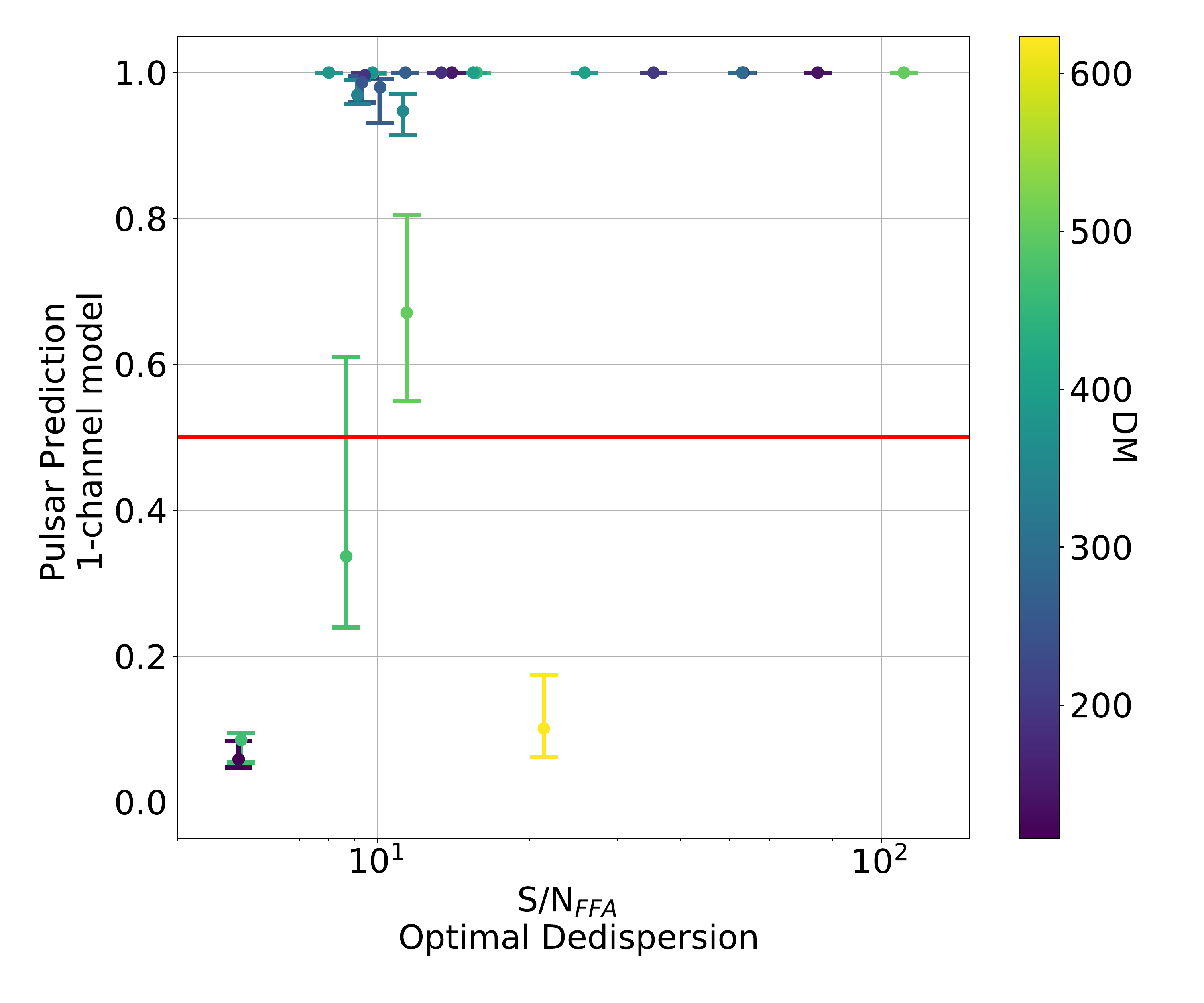}
 \caption{Comparison of the S/N$_{FFA}$ of the optimally dedispersed time series and the Pulsar Prediction of the one-channel model. The structure of the plot is the same as in Figure \ref{fig:model1_class} but with the one-channel model instead of the two-channel model. Similar to the performance of the two-channel model the weakest pulsars are not classified correctly but this model also has less accuracy on high-DM pulsars.}
 \label{fig:model2_class}
\end{figure}

The ensemble prediction of the five networks that were trained with each model applied on five partly overlapping segments of each observation are shown in Table \ref{tab:psr}. For the two-channel and one-channel models the same data are illustrated in figures \ref{fig:model1_class} and \ref{fig:model2_class}. The ensemble of the five trained networks of the two-channel model is able to correctly identify 22 out of 24 pulsars. Most of the pulsars above a S/N of 7 in the optimally dedispersed output are confidently classified as pulsars. In agreement with the results of Figure \ref{fig:ffa_model12} the one-channel model does have more problems at detecting high-DM pulsars than the two-channel model.

\begin{figure}
 \includegraphics[width=\columnwidth]{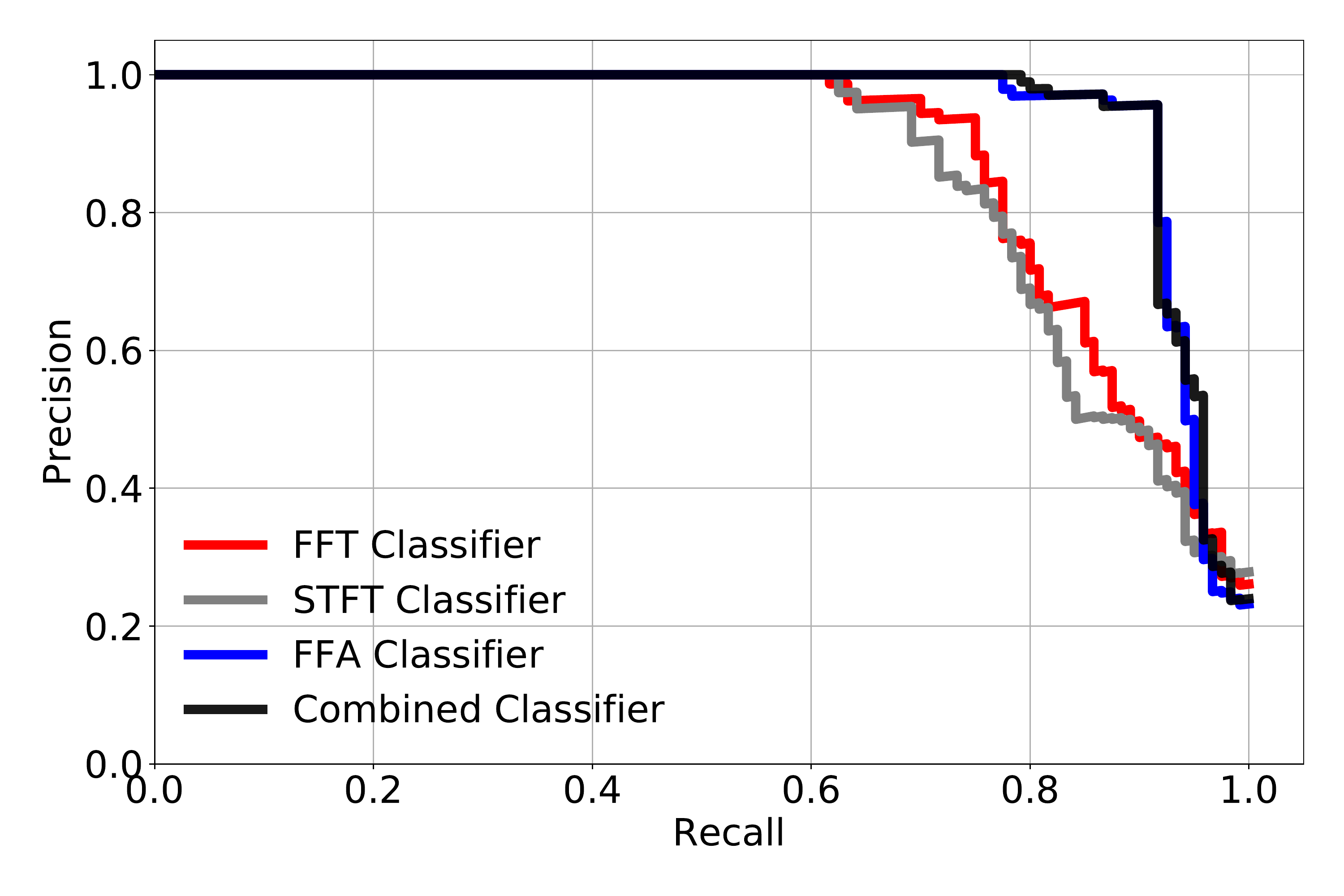}
 \caption{Precision-recall curves for the individual classifiers and the combined classifier in the two-channel model. For each observation in the test set five partly overlapping segments were tested.}
 \label{fig:prec_recall}
\end{figure}

Figure \ref{fig:prec_recall} shows the performance of the individual classifiers on the test set using the two-channel model. Clearly the overall prediction is dominated by the FFA classifier.
This shows that the used FFA method provides an easier input for classification than the FFT and STFT methods. 
This can also be seen in the weights of the individual classifiers. In the case of the two-channel model the mean weight of the FFT classifier is 0.25, the weight of the STFT classifier is 0.17 and the weight of FFA classifier is 0.75. The learning rates used in the third step are high enough that classifiers that are not useful for the classification of the training data can have a weight of close to 0 in the final classification. Since the weights of the two weaker classifiers are not completely scaled down they still are useful in the training set. 

While the FFT and the STFT classifiers perform worse they still provided useful gradients to the dedispersion network as seen in the comparison with the two-step model in Figure \ref{fig:mcc_dev}. 


When using the ensemble of networks in the two-channel model as is used in Figure \ref{fig:prec_recall_singles} only one out of 81 files in the set which do not contain a known pulsar are classified as containing a pulsar. In the case of this particular file the FFA shows a clear periodicity which is a harmonic of the period of the electrical grid. 

\begin{figure}
 \includegraphics[width=\columnwidth]{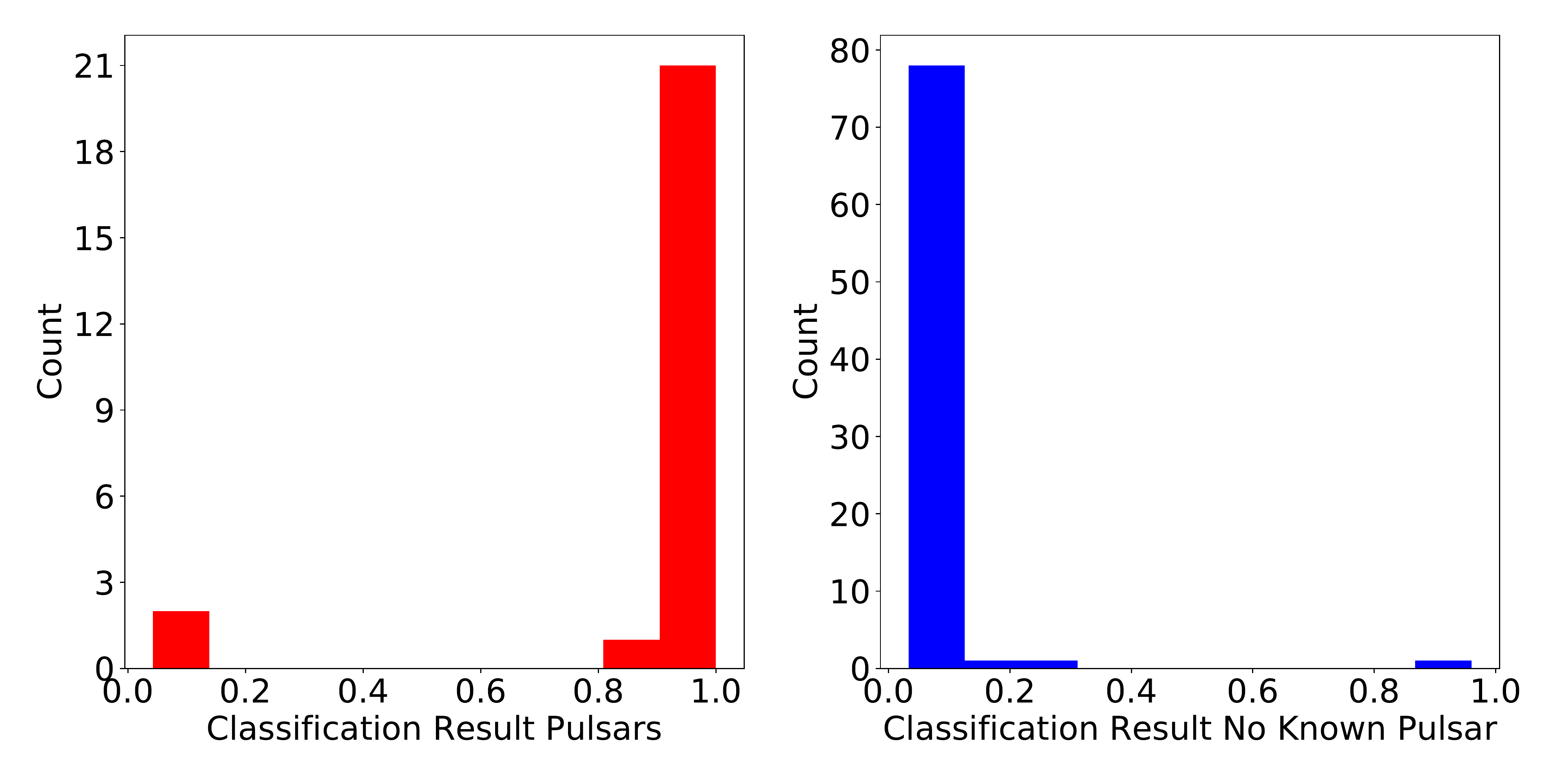}
 \caption{Histograms of the pulsar predictions of the 2-channel model for the test set. Both the pulsar predictions for the observations containing pulsars (left) and for the observations containing no known pulsars (right) show that the model either strongly predicts a pulsar or strongly predicts no pulsar.}
 \label{fig:model2_hist}
\end{figure}

Figure \ref{fig:model2_hist} shows what kind of pulsar predictions are typical of the 2-channel model.
The model either makes a strong prediction that there is a pulsar in the data or that there is no pulsar in the data.
One of the files not containing a known pulsar is confidently misclassified due to an RFI signal which was not completely removed be the neural network dedispersion. In the other observations only pulsar signals result in a positive classification. Two faint pulsars are missed. 
These two pulsar when optimally dedispersed only show a S/N$_{FFA}$ of 5.3 and 5.4 which means they would not typically be folded in a normal survey.

%% file: rfi.tex
\label{sec:rfi}
\begin{figure*}
 \includegraphics[width=\textwidth]{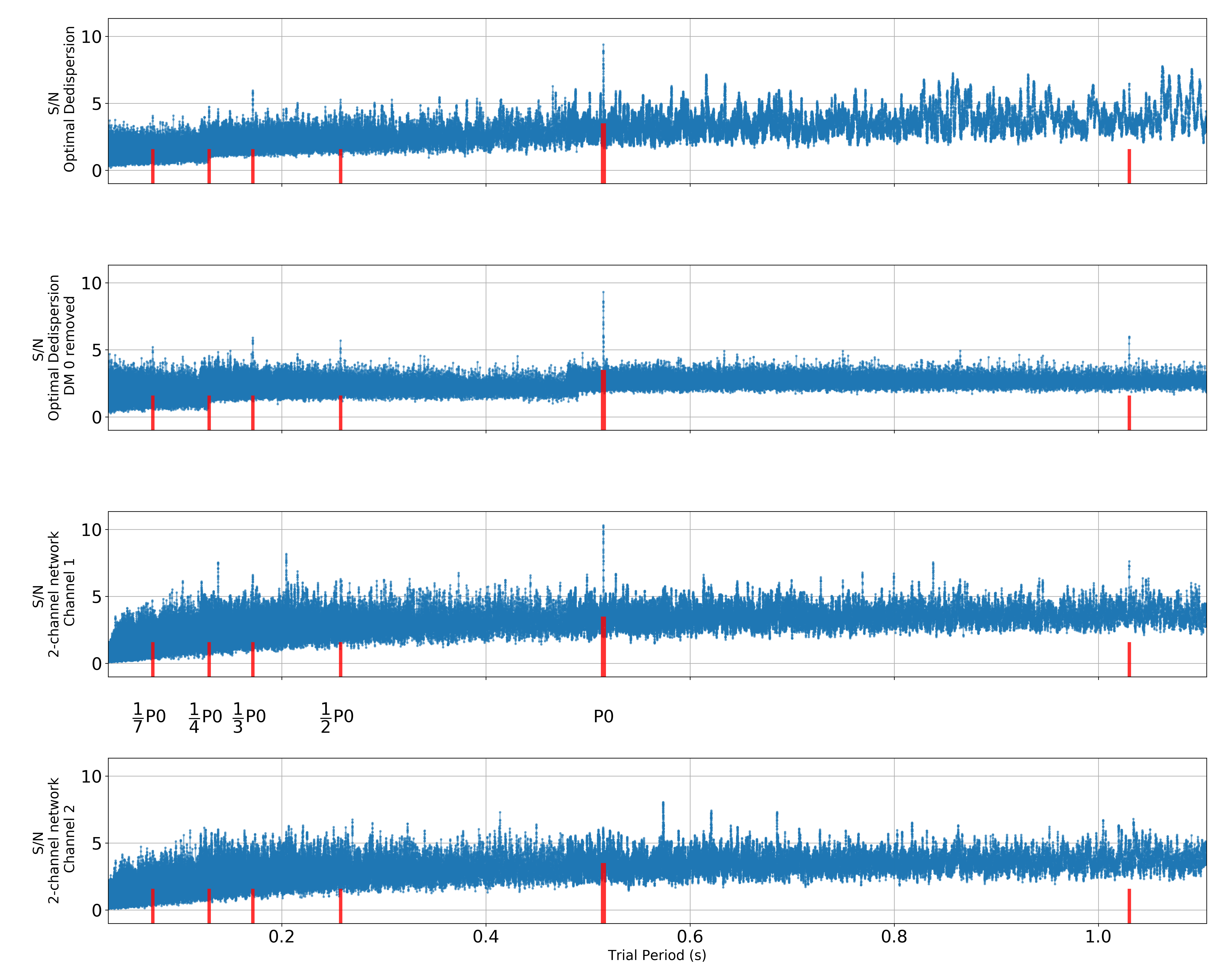}
 \caption{Comparison of the FFA of the observation of PSR J2007+3120 using optimal dedispersion and using one of the networks of the 2-channel model. First row: Optimal dedispersion. Second Row: Optimal dedispersion after removing the mean of each time step. Third row: Output channel 1 of the neural network. Fourth row: Output channel 2 of the neural network.
The pulsar has  a period of 0.608 ms and a DM of 192. The red lines show the pulse period and various fractions of it which also create peaks in the FFA}
 \label{fig:ffa_rfi_17}
\end{figure*}

In order to understand the influence of RFI on the intermediate time series and the final classification we take a closer look at PSR J2007+3120. In Figure \ref{fig:ffa_rfi_17} the FFA of this pulsar is shown using different input values. Removing the mean of each time step very effectively removes the noise in the FFA which is created by RFI. But even though RFI is suppressed the S/N of the actual pulsar is not increased. Using our model we observe that the S/N of the pulsar signal is increased and the noise in the FFA is decreased compared to only using optimal dedispersion. The noise in the FFA of the network output is still higher than in the case where the mean of each time step was subtracted. The peak at one seventh of the pulse period is much clearer in the output of the network than in the output of the optimal dedispersion.
Due to its low DM of 192 the pulsar only creates a strong signal in one of the output channels.

The observations in the FFA show that the created neural network structure is able to mitigate the effects of RFI. In this case we are only comparing to the simple baseline of removing RFI by removing the mean of each time step but it clearly shows that a network that is mostly trained by a classification loss is able to suppress RFI signals (and boost pulsar signals). 

In our case the RFI mitigation is done as part of our convolutional neural network. In the case of other data sets additional layers at the beginning of the network which employ an RFI masking approach similar to many RFI mitigation techniques could also be employed. The parameters of these layers could also be trained using the classification loss as long as these layers are differentiable.

%% file: application.tex
In this section we discuss the applicability of the neural network to pulsar surveys and possible future developments.

\paragraph*{Twofold Applicability to Pulsar Surveys}
As we have shown in sections \ref{sec:ffa_perf} and \ref{sec:perf_real} neural networks can be used to both dedisperse pulses of unknown DM and detect pulsars in the resulting time series. This allows the proposed neural network to enhance existing pulsar survey pipelines in two ways. The dedispersion can provide an additional time series that can be analysed with existing techniques to create multiple candidates or the whole network can be used to classify the observations.

\paragraph*{Finding Weaker Signals}
During the development of the classifiers we decided to use processing steps that explicitly use the periodicity of the pulses like the FFT and FFA instead of a purely convolutional neural network. The classification scheme currently only classifies based on the most significant signal of these processing steps. This means that this classification will not properly find pulsars when the pulsar signal is not stronger than the strongest noise-induced signals. How strong these noise-induced signals are heavily depends on the performance of the RFI mitigation of the system and the properties of the previous processing steps. 

Trading recall for precision is possible by training the neural network to a higher noise level or creating pulsar candidates from the output of the dedispersion network instead of classifying the whole observation.

Instead of classifying the whole observation one could also use the presented model by replacing the global max pooling operation in the classifier with an operation that outputs multiple candidates. This way the network can output real pulsar candidates even if there are RFI candidates which are stronger.

\paragraph*{Accelerated Pulsars}
In this work we only tested our methods on slow and unaccelerated pulsars but the network could also be adapted to pulsars with shorter or modulated pulse periods. Sensitivity to faster pulsars would be reached by reducing the initial downsampling of the filterbank files which will increase the memory and processing requirements of the neural networks but does not change anything fundamental of the required neural network architecture.

Since the convolutional dedispersion network only has a relatively small receptive field the performance of the dedispersion should not be affected by the presence of acceleration in the pulsar signals. The introduction of a classifier based on some sort of acceleration search (such as the algorithm used in \citet{2013MNRAS.431..292E} for example) could provide sensitivity to accelerated signals. If such an acceleration-based classifier performs well when trying to classify the complete observation remains to be tested.

Training a network to be sensitive to accelerated data creates the need to provide an accelerated training signal. This can be achieved by simulating accelerating pulsars or introducing acceleration via data augmentation. Under the assumption that the dedispersion network performs similarly well with the same weights on accelerated and unaccelerated signals, time-domain resampling of the dedispersed time series could be enough to train the acceleration-search based classifiers.

\paragraph*{Higher Data Volume}
This work explores the capability of neural networks to classify pulsar observations that have been heavily downsampled. We can train the chosen network on a GPU with 32 GB memory with a batch size of 15. After all preprocessing steps the input array to the network had a size of $14\times400,000$. Processing data with a larger volume will easily require more memory than is available in current GPUs, which hinders the end-to-end training approach. 

Processing data of a higher volume could still be done by operating on smaller time segments. The dedispersion network only has a relatively small receptive field when compared to the whole observation which means it can be expected to perform similarly well when it is applied to smaller chunks of the input. The output when processing these chunks could later be combined for the classification steps.

In going to observations with larger bandwidth the increased complexity of the shape of pulsar signal can increase the difficulty of dedispersing with a neural network. When looking at data with a larger bandwidth at a similar center frequency the pulsar pulses will be spread over a larger number of time steps. A neural net able to detect these signals with a higher temporal spread may require more channels in the neural network than in this work. 

When working with data with a higher resolution the reduced S/N per pixel can pose a problem for the dedispersion. It could be that in this case to building a neural network that is able to detect pulsars looking at the DM-time plane as in \citet{2018AJ....156..256C} might produce better results than looking at the original frequency-time plane.

\paragraph*{Label-Switching}

In this work we manually filtered out known pulsars from our training set. Using real pulsars in the training set can also be done but has not been done in this study due to the low number of pulsars.
The manual filtering of real pulsars can be circumvented by performing label-switching of the training set based on the classification result of the neural network.

This can be achieved by first assuming that every observation does not contain a pulsar and do the usual training using simulated pulsars. In an actual survey we usually have more observations containing no pulsar than containing a pulsar which means our network should still be able to train decently even when some of the labels are false. By looking at the classification results of the training set we can identify the strongest real pulsars in the training set and change the labels of these observations. 

With increasingly weaker simulated pulsars and more relabeled pulsar observations the network will be able to identify ever fainter pulsars in the training set.
During this whole process one has to prevent that files containing RFI are accidentally labelled as containing a pulsar.

%% file: conclusion.tex
In this work we show the capability of a convolutional neural network to find pulsars directly in filterbank data while showing a small number of false positive and false negative classifications.
The results of Section \ref{sec:ffa_perf} show that the dedispersing and denoising capabilities of the neural network produce a very useful output for the application of survey search techniques like the FFA. We show that in this data set the peaks of the pulsars in the FFA are stronger when using our network than when using normal dedispersion. While normal pulsar search pipelines have to carry out a range of DM trials we show that our network produces a competitive output with only one or two dedispersed time series.
We demonstrate that we can reach a good performance on real pulsars while only using
simulated pulsars during training. 
The RFI suppression of the network is trained by adding the simulated pulses to real observations.

On early data of the PALFA survey we are able to get a prediction for our 105 test observations which span 469 minutes in less than 2 minutes.
The processing time depends on the used computing hardware and the input data volume.
The network can be adapted to data from other surveys but will have to be retrained due to
differences in the observation properties and RFI characteristics.
Retraining the network on new data may take time on the order of days but fine tuning the network
architecture and training procedure to reach a good sensitivity may take several iterations 
and is therefore significantly more time-consuming than mere retraining is..
The code we used to obtain the simulated data and train our network is publicly available and can be applied to other data sets.
We are currently in the process of developing additional refinements such as label-switching (i.e. semi-supervised learning)
which allows training when it is not clear if pulsars are present in the observations used as training noise;
and a candidate-based output for our neural network that is able to identify multiple signals in one observation  instead of classifying the observation as a whole.

We demonstrate how a dedispersion neural network can be trained.
In contrast to the neural network based work reviewed in Section \ref{sec:astroml} which either uses a short segment of the raw observation or a dedispersed input to detect FRBs, we use a long segment of an observation and produce a classification result by first producing a dedispersed time series.
The creation of the dedispersed time series is trained by a combination of a reconstruction loss, which directly trains the network to create dedispersed pulses, and a classification loss.
We show that the classification loss of an FFT-based classifier can provide useful gradients for the training of the dedispersion network which creates this time series. 
Our current neural network architecture is primarily based on convolutions. Recent advances in sequence modelling that focus on attention and transformer-based models have shown superior results both in terms of accuracy and throughput \citep{attention2017}. We are working towards applying similar techniques to our data set that will likely boost the performance while demanding lower hardware requirements.

In order for our method to become a standard method in pulsar survey analysis 
testing the network in a blind survey context is necessary in order
to understand how well our model would perform
in a real survey context
without being optimised for a small set of test pulsars.
For that we are currently testing an adaptation of our network on data of the 
Parkes Multibeam Pulsar Survey 
to get a further understanding of how well our model performs in a survey with longer observations, larger bandwidth, lower signal-to-noise ratio per pixel 
and a large number of real pulsars.
Applying our model to whole observations of state-of-the-art surveys may currently 
be prohibited by memory demands but could be alleviated by future generations 
of GPUs with higher memory,  or by applying our dedispersion network on time- or frequency-subsets of the data, which can be recombined before classification.

%% file: acknowledgements.tex
We thank Sotiris Sanidas and Ewan Barr for the introduction into pulsar survey techniques. 
We thank Ewan Barr also for giving comments to improve the paper.
We thank the anonymous referee for the useful comments that improved the paper.

The Quadro P6000 used for prototyping the neural network was donated by the
NVIDIA Corporation.
JPWV acknowledges support by the Deutsche Forschungsgemeinschaft (DFG) through the Heisenberg programme (Project No. 433075039).
Further developments and the final training of the neural networks has been done on the Bielefeld GPU cluster.
Part of this research has made use of the EPN Database of Pulsar Profiles maintained by the University of Manchester, available at: \url{http://www.epta.eu.org/epndb/ .}
This work is based on observations of the Arecibo Observatory. The Arecibo Observatory is a facility of the National Science Foundation operated under cooperative agreement by the University of Central Florida and in alliance with Universidad Ana G. Mendez, and Yang Enterprises, Inc.
We thank the developers of \textsc{pytorch} and all other software packages used in this work.
We acknowledge financial support by the German Federal Ministry of  Education  and  Research  (BMBF)  under  grant  05A17PB1  (Verbundprojekt D-MeerKAT).